\documentclass[useAMS,usenatbib,usegraphicx,onecolumn]{mn2e}

\title[Godunov smoothed  particle magnetohydrodynamics 
]{Smoothed  particle magnetohydrodynamics with 
a Riemann solver and the method of characteristics}

\author[Kazunari Iwasaki and Shu-ichiro Inutsuka]{Kazunari Iwasaki$^{1}$\thanks{E-mail:
iwasaki@nagoya-u.jp, inutsuka@nagoya-u.jp} and Shu-ichiro Inutsuka$^{1}$\footnotemark[1]
\\
$^{1}$Department of Physics, Nagoya University, Furo-cho, 
Chikusa-ku, Nagoya, Aichi, 464-8602, Japan}
\begin{document}

\date{Accepted Received}

\pagerange{\pageref{firstpage}--\pageref{lastpage}} \pubyear{2002}

\maketitle

\label{firstpage}

\begin{abstract}
In this paper, we develop a new method for magnetohydrodynamics (MHD) using 
smoothed particle hydrodynamics (SPH). To describe MHD shocks accurately, 
the Godunov method is applied to SPH instead of artificial dissipation terms.
In the interaction between particles,
we solve a non-linear Riemann problem with magnetic pressure for compressive waves  
and apply the method of characteristics for Alfv{\'e}n waves.
An extensive series of MHD test calculations is performed.  
In all test calculations, we compare the results of our SPH code with
those of a finite-volume method with an approximate Riemann solver, and 
confirm excellent agreement.
\end{abstract}


\section{Introduction}
It is well known that magnetic fields play an important role
in various astrophysical phenomena, such as the formation of stars and planets, and 
high energy astrophysics. 
Gas dynamics with a magnetic field is well described by magnetohydrodynamics (MHD).
The MHD equations are too complicated to solve analytically except for special cases,
therefore to understand physical phenomena involving magnetic fields, 
numerical simulations are indispensable and powerful tools. 
A numerical technique to solve the MHD equations has been developed
successfully in the form of the finite-volume methods by many authors.

Smoothed particle hydrodynamics (SPH) is a fully Lagrangian particle method \citep{L77,GM77}.
This Lagrangian nature has major advantages in problems that have a large dynamic range
in spatial scale, such as
the formation of large scale structures, galaxies, stars, and planets.
Several authors have tried to apply the SPH method to MHD problems.
In this paper, we call SPH for MHD ``smoothed particle magnetohydrodynamics'' (SPMHD).
\citet{PM04a,PM04b} have developed an one-dimensional SPMHD scheme.
To describe shock waves, 
they used artificial dissipation terms proposed by \citet{M97} based
on an analogy with Riemann solutions of compressible gas dynamics.
As the signal velocity, they adopted the speed of the fast wave,
which is analogous to the sound wave in hydrodynamics (HD).
Their SPMHD has been shown to give good results on a wide range of standard one-dimensional 
problems used in recent finite-volume MHD schemes.
\citet[][hereafter PM05]{PM05} have developed a multi-dimensional SPMHD scheme based on the 
above one-dimensional one.
Alternatively, \citet{Betal01} and \citet{Betal06} have implemented an SPMHD scheme using 
a regularization of the underlying particle distribution and artificial viscosity.
Recently, \citet{DS09} have implemented MHD in the cosmological SPH code {\it GADGET}
\citep{Setal01,S05} mainly based on PM05.
Broad discussions of the SPH and SPMHD are found in reviews 
by \citet{M92}, \citet{S10a}, and \citet{P10}. 

The SPMHD in PM05 can capture fast shocks accurately.
However, since they use the fast wave speed as the signal velocity in the artificial 
viscosity and resistivity terms, Alfv{\'e}n waves become dissipative.
In the finite-volume method, it is well known that 
Alfv{\'e}n waves cannot be described accurately 
if the characteristics of Alfv{\'e}n waves 
are not taken into account in the caluculation of the numerical flux \citep[e.g.,][]{SN92}.
This is also the case in SPMHD. 
In this paper, we apply the Godunov method to SPMHD.
The Godunov method was originally developed in the 
finite-volume method \citep{G59,vL79}.
Unlike artificial viscosity, the Godunov method can, in principle, take into account 
the minimum and sufficient amount of dissipation without any free parameters.  
In HD, the application of the Godunov method to SPH has been 
carrried out by \citet[][hereafter I02]{I02}.
In MHD, we can also consider the general Riemann problem (RP) with
arbitrary directions of velocity and magnetic field in both sides.
However, it is computationally expensive and complex to solve the RP because the 
MHD equations have seven characteristics and non-hyperbolicity. 
Recently, \citet{GN11} have approximate Riemann solver HLLC into a variant particle method
\citep{TS94,L05}.
Besides the SPH method, \citet{PBS11} implemented MHD with the approximate Riemann solver HLLD
\citep{MK05} in unstructured, moving-mesh code AREPO code \citep{S10b}.
In this paper, we use a simplified approach proposed 
for the finite-volume method by \citet{Setal99}.
In this method, compressible and incompressible parts of the MHD equations are completely 
divided. 
The former is calculated by a non-linear Riemann solver with magnetic pressure, and 
the latter is calculated by the method of characteristics (MOC) proposed by \citet{SN92}.

The paper is organized in the following ways.
In Section \ref{sec:spheq}, we derive the SPMHD equations from the basic equations of MHD.
In Section \ref{sec:implementation}, we describe the implementation of the derived SPMHD equations.
Various test calculations are demonstrated in Section \ref{sec:test}.
This paper is summarized in Section \ref{sec:summary}.

\section{SPMHD equations}\label{sec:spheq}
The basic equations of MHD can be written as
\begin{equation}
    \frac{d}{dt} 
    \left(
    \begin{array}{c}
            1/\rho \\
            v^\mu\\
            E   \\
    \end{array}
    \right)
    =  \frac{1}{\rho}\nabla^\nu
    \left(
    \begin{array}{c}
            v^\nu \\
            T^{\mu\nu} \\
            T^{\mu\nu}v^\mu   \\
    \end{array}
    \right),
    \label{basic eq}
\end{equation}
\begin{equation}
        \frac{d}{dt}\left( \frac{B^\mu}{\rho} \right)
        = \frac{B^\nu}{\rho}\nabla^\nu v^\mu,
        \label{induc}
\end{equation}
where $T^{\mu\nu}$ is the stress tensor,
\begin{equation}
        T^{\mu\nu} = -\left( P+\frac{\bmath{B}^2}{2} \right)\delta^{\mu\nu} 
        + B^\mu B^\nu,
\end{equation}
the specific total energy is given by
\begin{equation}
        E = \frac{1}{2} \bmath{v}^2 + e + \frac{\bmath{B}^2}{2\rho},
\end{equation}
$e=P/[(\gamma-1)\rho]$ is the specific internal energy,
$d/dt=\partial/\partial t + \bmath{v}\cdot\bmath{\nabla}$ is the Lagrangian time derivative,
and we have chosen units so that factor of $\mu_0$ does not appear in the equations, 
where $\mu_0$ is the magnetic permeability.

In the SPH method, density field is expressed as 
\begin{equation}
        \rho(\bmath{x}) = \sum_j m_j W(\bmath{x}-\bmath{x}_j,h(\bmath{x})),
        \label{sph den}
\end{equation}
where subscripts denote particle labels, $m_j$ is the mass of the $j$-th particle,
$W(\bmath{x},h)$ is a kernel function, and $h$ is the smoothing length that is assumed to 
depend on $\bmath{x}$.
There are many choices of kernel functions.
In the Godunov SPH (GSPH) schemes, I02 adopted the following Gaussian kernel,
\begin{equation}
        W(\bmath{x},h) = \left( \frac{1}{\sqrt{\pi}h} \right)^d e^{-(\bmath{x}/h)^2},
        \label{kernel}
\end{equation}
where $d$ is the number of dimension. 
This is because the Gaussian kernel makes the formulation of the GSPH simpler than 
other kernel functions.
In addition, the results with Gaussian kernel seem to be better than those with 
the cubic spline kernel in test calculations shown in section \ref{sec:test}.
In this paper, we follow the choice in I02.

\subsection{Equation of motion}
We define the equation for the evolution of the particle positions as 
\begin{equation}
        \ddot{\bmath{x}}_i \equiv 
     \int d^3 x\frac{d \bmath{v}}{d t} W_i(\bmath{x}),
        \label{eom1}
\end{equation}
where $W_i(\bmath{x})\equiv W[\bmath{x}-\bmath{x}_i,h(\bmath{x})]$. 
Substituting equation (\ref{basic eq}) into equation (\ref{eom1}), one obtains
\begin{equation}
  \ddot{x}_i^\mu =  
  \int d^3x \frac{1}{\rho}\left(\nabla^\nu T^{\mu\nu}\right) W_i(\bmath{x})
  = \int d^3 x \nabla^\nu\left(T^{\mu\nu}\frac{W_i(\bmath{x})}{\rho} \right)
  - \int d^3x T^{\mu\nu}\nabla^\nu\left(\frac{W_i(\bmath{x})}{\rho} \right),
        \label{eom2}
\end{equation}
where we integrate by part.
The first term on the right-hand side in equation (\ref{eom2}) 
becomes the surface integral by the Gauss theorem, and vanishes when $|\bmath{x}|\rightarrow\infty$.
With equation (\ref{sph den}), the last factor of equation (\ref{eom2}) becomes
\begin{equation}
   \nabla^\nu\left(\frac{W_i(\bmath{x})}{\rho} \right) = 
   \frac{1}{\rho^2}\sum_jm_j \left( \nabla^\nu W_i(\bmath{x})W_j(\bmath{x}) 
    - W_i(\bmath{x}) \nabla^\nu W_j(\bmath{x})\right). 
        \label{eom3}
\end{equation}
Therefore, we can obtain the equation for the evolution of the 
particle positions,
\begin{equation}
  \ddot{x}_i^\mu =  
 - \sum_j m_j \int d^3x \frac{T^{\mu\nu}}{\rho^2} \left\{ \nabla^\nu W_i(\bmath{x})W_j(\bmath{x}) 
  - W_i(\bmath{x}) \nabla^\nu W_j(\bmath{x})\right\}.
        \label{eom sph}
\end{equation}

\subsection{Total energy equation}
We define the equation for the evolution of the particle energy as 
\begin{equation}
  \dot{E_i} \equiv 
     \int d^3 x\frac{d E_i}{d t} W_i(\bmath{x}).
      \label{eoe1}
\end{equation}
Substituting equation (\ref{basic eq}) into equation (\ref{eoe1}), one obtains 
\begin{equation}
  \dot{E}_i =  
  \int d^3x \frac{1}{\rho}\left(\nabla^\mu T^{\mu\nu} v^\nu\right) W_i(\bmath{x}) 
  =- \int d^3x T^{\mu\nu} v^\nu\nabla^\mu\left(\frac{W_i(\bmath{x})}{\rho} \right).
    \label{eoe2}
\end{equation}
Using equation (\ref{eom3}), we can obtain the energy equation of SPH particles,
\begin{equation}
  \dot{E_i} =  
  - \sum_j m_j \int d^3x 
  \frac{T^{\mu\nu}v^\nu}{\rho^2} 
\left\{ \nabla^\nu W_i(\bmath{x})W_j(\bmath{x}) 
  - W_i(\bmath{x}) \nabla^\nu W_j(\bmath{x})\right\}
        \label{eoe sph}
\end{equation}

\subsection{Induction equation}
The $i$-th SPH particle is assigned its own magnetic field, $\bmath{B}_i$.
The equation for the evolution of the magnetic field is defined as
\begin{equation}
        \frac{d}{dt} \left( \frac{\bmath{B}}{\rho} \right)_i
        \equiv \int d^3x
        \frac{d}{dt} \left( \frac{\bmath{B}}{\rho} \right) W_i(\bmath{x}) .
        \label{induc1}
\end{equation}
Substituting equation (\ref{induc}) into equation (\ref{induc1}), one obtains 
\begin{equation}
    \frac{d}{dt} \left( \frac{B^\mu}{\rho} \right)_i = \int   d^3x
         \frac{B^\nu}{\rho}\nabla^\nu v^\mu W_i(\bmath{x})
    \label{induc2}
\end{equation}
The right-hand side of equation (\ref{induc2}) can be transformed into the following expression:
\begin{equation}
  \int  d^3x\frac{B^\nu}{\rho}\nabla^\nu v^\mu W_i(\bmath{x}) = 
  \int  d^3x \frac{\nabla^\nu(B^\nu v^\mu)}{\rho} W_i(\bmath{x})
- \int d^3x v^\mu\frac{\nabla^\nu B^\nu}{\rho} W_i(\bmath{x}) 
    \label{induc3}
\end{equation}
We approximate the last term on the right-hand side of equation (\ref{induc2}) as follows:
\begin{equation}
 \int  v^\mu\frac{\nabla^\nu B^\nu}{\rho} W_i(\bmath{x}) d^3x
 = \dot{x}_i^\mu \int  \frac{\nabla^\nu B^\nu}{\rho} W_i(\bmath{x}) d^3x+ O(h^2).
    \label{induc3}
\end{equation}
Using equations (\ref{induc2}) and (\ref{induc3}) and integrating by parts, 
one obtains the following equation for the evolution of the magnetic field: 
\begin{equation}
  \frac{d}{dt} \left( \frac{B^\mu}{\rho} \right)_i = 
  -\sum_j m_j \int d^3x
  \frac{B^{\nu}}{\rho^2}\left( v^\mu - \dot{x}_i^\mu \right) 
\left\{ \nabla^\nu W_i(\bmath{x})W_j(\bmath{x}) 
  - W_i(\bmath{x}) \nabla^\nu W_j(\bmath{x})\right\}.
        \label{induc sph}
\end{equation}

\section{Implementation}\label{sec:implementation}
\subsection{Convolution}
We define $s$-axis as being along the vector $\bmath{x}_i-\bmath{x}_j$. 
The unit vector in the $s$-direction is $\bmath{n}=(\bmath{x}_i - \bmath{x}_j)/|\bmath{x}_i - \bmath{x}_j|$.
The distance between $i$- and $j$-th particles is $\Delta s_{ij}=s_i-s_j$ 
where $s_i$ and $s_j$ are the coordinates of 
the $i$- and $j$-th particles on the $s$-axis, respectively.
We need to perform the volume integral in equations (\ref{eom sph}), (\ref{eoe sph}), and 
(\ref{induc sph}). 
If the smoothing length is spatially constant, 
the volume integral can be done analytically by interpolating physical variables \citep{I02}.
In a similar way, one can derive the GSPMHD equations as follows:
\begin{equation}
        \frac{d}{dt} \left(
        \begin{array}{c}
                  \dot{x}_i^\mu \\
                  E_i \\
                  (B^\mu/\rho)_i
        \end{array}
        \right)
        = \sum_j m_j 
        F_{ij}
        \left(
        \begin{array}{c}
           (T^{\mu\nu})^* n^\nu \\
           n^\mu(T^{\mu\nu}v^\nu)^* \\
           (B^\nu)^* n^\nu \left\{ (v^\mu)^* - \dot{x}_i^{*\mu} \right\} 
        \end{array}
        \right),
        \label{sph eqs}
\end{equation}
where
\begin{equation}
      F_{ij} = 2V_{ij}^2(h)\frac{\partial W(\Delta s_{ij},h)}{\partial s_i},
\end{equation}
$\left( T^{\mu\nu} \right)^*$, $\bmath{v}^*$, and $\bmath{B}^*$ are the values at $s=s_{ij}^*$ 
\citep[see][]{I02}, and 
$\dot{\bmath{x}}_i^{*} = \bmath{v}_i + \ddot{\bmath{x}}_i\Delta t /2$ is the 
time-centered velocity of the $i$-th particle.
The quantity $V_{ij}^2(h)$ is obtained by the following integration
\begin{equation}
    \int \rho^{-2} W_i(\bmath{x}) W_j(\bmath{x}) 
    d^3 x = V_{ij}^2(h)W(\bmath{x}_i-\bmath{x}_j,\sqrt{2}h),
\end{equation}
where I02 used linear and cubic interpolations of $\rho^{-1}(\bmath{x})$.
The detailed expression of $V_{ij}^2(h)$ is found in I02.
However, for the case with variable smoothing length, 
the volume integral cannot 
be easily performed analytically.
I02 proposed a simple approximation in which
he used $h_i$ for the half of the integration space 
that includes $x_i$ and $h_j$ for the other half.
In this case, $F_{ij}$ becomes
\begin{equation}
        F_{ij} = V_{ij}^2(h_i)\frac{\partial W(\Delta s_{ij},\sqrt{2}h_i)}
        {\partial s_i}
        + V_{ij}^2(h_j)\frac{\partial W(\Delta s_{ij},\sqrt{2}h_j)}{\partial s_i}.
        \label{Fij I02}
\end{equation}
This formulation can capture contact discontinuities accurately \citep{CIN10,Metal11}.
In this paper, just for simplicity, we use the following crude approximation   
in the volume integral of equations (\ref{eom sph}), (\ref{eoe sph}), and (\ref{induc sph}): 
\begin{equation}
        \bmath{\nabla} W_i(\bmath{x})W_j(\bmath{x}) - W_i \bmath{\nabla} W_j(\bmath{x}) 
        \simeq \bmath{\nabla} W_i(\bmath{x})\delta(\bmath{x}-\bmath{x}_j) 
        - \delta(\bmath{x}-\bmath{x}_i)\bmath{\nabla} W_j(\bmath{x}).
\end{equation}
Using this, one can get 
\begin{equation}
        F_{ij} = \left( \frac{1}{\rho_i^2} + \frac{1}{\rho_j^2} \right) \frac{\partial W(\Delta s_{ij},\bar{h}_{ij})}
        {\partial s_i},
        \label{Fij monaghan}
\end{equation}
where $\bar{h}_{ij}$ is an average of $h_i$ and $h_j$.
This paper adopts  $\bar{h}_{ij}=(h_i + h_j)/2$.

\subsection{The usage of the Riemann solver}\label{sec:use riemann solver}
The equation (\ref{sph eqs}) do not include a dissipative process, which is required to describe shock waves.
The Godunov method uses the exact Riemann solver to include the minimum and sufficient amount of 
dissipation into the scheme.
In the finite-volume method, the result of the Riemann problem at cell interfaces is used in the calculation
of numerical flux.
In the GSPH in I02, the values $P^*$ and $(Pv)^*$ in equations (58) and (59) in his paper 
are replaced by the results of the RP between the $i$-th and the $j$-th particles.
In the same way, $\left( T^{\mu\nu} \right)^*$ and $\left( T^{\mu\nu}v^\nu \right)^*$
are replaced by the results of the RP between the $i$-th and the $j$-th particles.
In equation (\ref{sph eqs}),
the projection of $T^{\mu\nu}$ on the $s$-axis is found as follows:
\begin{equation}
        T^{\mu\nu}n^\nu = -\left( P_\mathrm{t} - \frac{B_\parallel^2}{2}\right)n^\mu
        + B_\parallel B_\perp^\mu,
        \label{projection}
\end{equation}
where $P_\mathrm{t}\equiv P+B_\perp^2/2$, and
component parallel (perpendicular) to $\bmath{n}$ is represented by using the subscript of 
$\parallel$ ($\perp$).
The first term on the right-hand side of equation (\ref{projection}) represents the compressive term 
working alone the $s$-axis.
In contrast, the second term represents the incompressible term working in the perpendicular 
direction.
In the compressible part, we use the result of the non-linear RP without $B_\parallel$, which contains 
fast shocks, fast rarefaction waves, and one contact discontinuity.
From the RP, one can obtain $P_\mathrm{t}^*$ and $ v_\parallel^*$.
The detailed description is shown in Appendix \ref{riemann}.
In the incompressible term, MOC is used \citep{SN92}.
From the MOC, one can obtain $\bmath{B}_\perp^*$ and $\bmath{v}_\perp^*$.
The detailed description is shown in Appendix \ref{moc}.

In the calculation of the RP and the MOC, the initial values on each side at $s_{ij}^*$ are required.
In this paper, we adopt $s_{ij}^*=0$ for simplicity. It is confirmed that the value of $s_{ij}^*$ 
does not affect the results.
To make a spatially second-order method, we 
consider the piecewise linear distribution of the physical variables. 
Using the gradients, the initial values on each side of a 
one-dimensional RP are the average values of each domain 
of dependence:
\begin{equation}
        \bmath{U}_\mathrm{R} = \bmath{U}_i - \frac{1}{2}\left(\frac{\partial \bmath{U}}{\partial s}\right)_i
        \left[ \Delta s_{ij} - C_{i}\Delta t\right]
\end{equation}
\begin{equation}
        \bmath{U}_\mathrm{L} = \bmath{U}_j + \frac{1}{2}\left(\frac{\partial \bmath{U}}{\partial s}\right)_j
        \left[ \Delta s_{ij} - C_{j}\Delta t\right]
\end{equation}
where $\bmath{U}=(\rho,\;P,\;\bmath{v},\;\bmath{B})$, and 
$C$ is a characteristic speed.
In the compressible RP, the speed of the fast wave $C=\sqrt{(\gamma P + \bmath{B}^2)/\rho}$ is adopted.
In the incompressible RP, the speed of the Alfv{\'e}n wave $C=|B_\parallel|/\sqrt{\rho}$ is adopted.

In finite-volume methods with higher spatial accuracy, we need to impose a monotonicity constraint on
the gradients of the physical variable to obtain a stable description of discontinuities.
This is the case in the GSPMHD scheme. The detailed description of the monotonicity constraint adopted in this paper is 
shown in Appendix \ref{app:mono}.

In actual calculations, we solve the following equation: 
\begin{equation}
        \frac{d}{dt} \left(
        \begin{array}{c}
                \dot{\bmath{x}}_i \\
                  E_i \\
                  (\bmath{B}/\rho)_i
        \end{array}
        \right)
        = \sum_j m_j 
        F_{ij}
        \left(
        \begin{array}{c}
                -\left( (P_\mathrm{t})_\mathrm{RP} - P_\mathrm{B,\parallel}\right)\bmath{n}
                + B_{\parallel}^*(\bmath{B}_\perp)_\mathrm{MOC}
           \\
           -\left( (P_\mathrm{t})_\mathrm{RP} - P_\mathrm{B,\parallel}\right)(v_{\parallel})_\mathrm{RP}
           + B_{\parallel}^* (\bmath{B}_\perp)_\mathrm{MOC} \cdot(\bmath{v}_\perp)_\mathrm{MOC}\\
           B_{\parallel}^* \left\{ ( v_\parallel )_\mathrm{RP} \bmath{n} + 
           \left(\bmath{v}_\perp\right)_\mathrm{MOC} - \dot{\bmath{x}}_i^{*} \right\} 
        \end{array}
        \right),
        \label{sph eqs1}
\end{equation}
where subscripts ``RP'' and ``MOC'' indicate values
evaluated using the RP and MOC, respectively, and 
$P_\mathrm{B,\parallel}\equiv\left( B_{\parallel,i}^2 + B_{\parallel,j}^2 \right)/4$ and
$B_\parallel^*=\left( B_{\parallel,i} + B_{\parallel,j} \right)/2$.

\subsection{Variable smoothing length}
In this paper, the variable smoothing length is used to obtain large dynamic ranges.
The smoothing length of the $i$-th particle is determined iteratively by 
\begin{equation}
        h_i = C_h \left( \frac{m_i}{\sum_j m_j W(\bmath{x}_i-\bmath{x}_j,h_i)} \right)^{1/d},
        \label{hi}
\end{equation}
where $C_h$ is a parameter. 
The density of the $i$-th particle is evaluated by 
\begin{equation}
        \rho_i = \sum_j m_jW(\bmath{x}_i-\bmath{x}_j,h_i).
        \label{deni}
\end{equation}
The Gaussian kernel is not truncated at a finite radius but has infinite range.
In practical calculations, we ignore the contribution from the $j$-th to $i$-th particles
if $|\bmath{x}_i - \bmath{x}_j|>3.1h$ because $\exp(-3.1^2)\sim 6.7\times10^{-5}$ 
is sufficiently small. The number of neighbours becomes $\sim6C_h$ in 1D, $\sim30C_h^2$ in 2D, and $\sim124C_h^3$ in 
3D schemes.
In this paper, we present the results of 2D test calculations and adopt $C_h=1.2$, indicating that the average 
neighbour number is $\sim 43$.

\subsection{Corrections for Avoiding Tensile Instability
due to Magnetic Force}\label{sec:corr}
From equation (\ref{projection}), one can see that 
the stress tensor can be negative when 
the plasma beta $\beta\equiv 2P/\bmath{B}^2$ is low. This causes unphysical clumping of SPH particles \citep{M92}.
This numerical instability is called ``tensile instability'' \citep{Setal95}.
The tensile instability arises from the fact that the SPH 
expression of $\bmath{\nabla}\cdot \bmath{B}$ is not completely zero.
In finite-volume methods, it is well known that 
nonzero $\bmath{\nabla}\cdot \bmath{B}$ produces an unphysical force along 
the magnetic field and causes large errors in the simulations when 
using conservative form \citep{BB80}. 
Many authors proposed methods for vanishing $\bmath{\nabla}\cdot\bmath{B}$, 
such as the projection method \citep{BB80}, the constrained transport \citep{EH88}, and so on.

In SPMHD, $\bmath{\nabla}\cdot\bmath{B}$ inevitably has some amount of numerical noise that causes the 
tensile instability to occur even if divergence-cleaning methods are used in the conservation form \citep{PM05}.
Therefore, several methods have been proposed to suppress the tensile instability.
\citet{PM85} proposed that 
the stress tensor make positive by subtracting an constant value from the stress tensor 
\citep[also see][]{PM05}.
As another approach, \citet{Betal01} suggested that the 
monopole source terms are explicitly 
subtracted from the equation of motion and the energy equation  
as follows:
\begin{equation}
        \frac{dv^\mu}{dt} = \frac{1}{\rho}\nabla^\nu T^{\mu\nu}
        - \frac{1}{\rho}B^\mu \bmath{\nabla}\cdot \bmath{B},
        \label{eom corr}
\end{equation}
and
\begin{equation}
        \frac{dE}{dt}  
        = \frac{1}{\rho}\nabla^\mu \left(T^{\mu\nu} v^\nu\right)
        - \frac{1}{\rho}\left(\bmath{B}\cdot \bmath{v}\right) \bmath{\nabla}\cdot \bmath{B}.
        \label{eoe corr}
\end{equation}
The left-hand side of equation (\ref{eom corr}) corresponds to 
the Lorentz force $((\bmath{\nabla} \times \bmath{B})\times\bmath{B})/\rho$.
This source terms significantly stabilize the tensile instability.
This formulation is the same as so-called 8-wave formulations proposed by \citet{Petal99} 
in the finite-volume method.
This formulation is numerically stable, and  
the value of $\bmath{\nabla}\cdot\bmath{B}$ keeps zero within truncation error without any cleaning methods 
for simple test problems.
However, in realistic 3D problems,
satisfaction of the divergence constraint is not guaranteed. 
Thus, we need to monitor the value of $\bmath{\nabla}\cdot\bmath{B}$ in actual calculations.
To derive the GSPMHD equations, we take the convolution of the source term in equation (\ref{eom corr}),
\begin{equation}
   - \int  \frac{1}{\rho}\bmath{B} \bmath{\nabla}\cdot \bmath{B} W_i(\bmath{x})d^3 x
     \simeq 
    - \bmath{B}_i\int  \bmath{\nabla}\cdot \bmath{B} \frac{W_i(\bmath{x})}{\rho}d^3 x
    = \bmath{B}_i\int \bmath{B}\cdot \bmath{\nabla} \left(\frac{W_i(\bmath{x})}{\rho}\right)d^3 x.
\end{equation}
Using equation (\ref{eom3}), one can get
\begin{equation}
        \ddot{x}_i^\mu =  \sum_j m_j 
  \left\{(T^{\mu\nu})^*n^\nu  - B_i^\mu B_\parallel^*\right\} 
  F_{ij},
        \label{eom sph2}
\end{equation}
In a similar way, the energy equation is also obtains as 
\begin{equation}
        \dot{E}_i^\mu = \sum_j m_j 
        \left\{(T^{\mu\nu}v^\nu)^*  
        - \bmath{B}_i\cdot \dot{\bmath{x}}_i^* B_\parallel^*
        \right\} 
        F_{ij}.
        \label{eoe sph2}
\end{equation}
The major disadvantage of this correction is violation of the momentum and the total energy conservation.
Actually, in all SPMHD schemes without the tensile instability, 
the conservations of the momentum and/or total energy are sacrificed.

\subsection{Divergence Error Estimate}
In SPMHD, there are many choices of $\bmath{\nabla}\cdot\bmath{B}$ estimate.
In this paper, we use the following expression:
\begin{equation}
        \left(\frac{1}{\rho} \bmath{\nabla}\cdot\bmath{B}\right)_i = \int \frac{1}{\rho}\bmath{\nabla}\cdot\bmath{B}
        W(\bmath{x}-\bmath{x}_i)d^3 x
        = \sum_j m_j B_\parallel^* F_{ij}.
        \label{divB}
\end{equation}
In order to estimate errors in the $\bmath{\nabla}\cdot\bmath{B}=0$ constraint, we monitor $\delta_\mathrm{B}$ 
defined as
\begin{equation}
 \delta_\mathrm{B}=\frac{1}{N_\mathrm{tot}}\sum_i \frac{h_i| \rho_i \left(\bmath{\nabla}\cdot\bmath{B}/\rho\right)_i|}{|\bmath{B}_i|},
   \label{deltaB}
\end{equation}
where $N_\mathrm{tot}$ is the total number of the SPH particles.
This estimate is the same as that in the correction terms in section \ref{sec:corr}.
Note that PM05 and \citet{Betal06} adopted different choice, 
\begin{equation}
        \left(\bmath{\nabla}\cdot\bmath{B}\right)_i = \frac{1}{\rho_i}\sum_jm_j\left( \bmath{B}_i - \bmath{B}_j \right)
        \cdot \bmath{\nabla}_i W(\bmath{x}_i-\bmath{x}_j,h_i).
        \label{divB1}
\end{equation}
The divergence error estimated in equation (\ref{divB1}) tends to be smaller than those in equation (\ref{divB}).

\section{Numerical Tests}\label{sec:test}
In this section, we show the results of test calculations in the two-dimensional 
GSPMHD method, starting with the convergence test.
\subsection{Convergence Test}
In this section, we check the accuracy of the GSPMHD.
A good problem for the convergence test is the propagation test of linear MHD waves 
in a uniform media. First, we define the unperturbed state.
A uniform and rest gas is considered ($\rho=1$, $P=1$, $\bmath{v}=0$).
The magnetic field is parallel to the $x$-direction, and 
its magnitude is $1/\sqrt{2}$.
The simulations are performed in the square domain, $x,y\in\left[ 0,\sqrt{2}
\right]$. 
The particles are uniformly spaced on a cubic lattice with sides parallel to the $x$- and $y$-axes.
A periodic boundary condition is imposed. 

We consider linear MHD waves having a wavenumber 
of $\bmath{k}=2\pi(1/\sqrt{2},1/\sqrt{2},0)$, indicating that 
the simulation domain contains two wavelengths.
It is well known that MHD linear waves consist of the fast, 
Alfv{\'e}n, and slow waves with phase velocities
given by 1.4, 0.5, and 0.46, respectively in this configuration.
We add the initial perturbation based on the eigenmode.
In the fast and slow modes, 
all fluctuations lie in the $(x,y)$ plane.
The amplitude of the density fluctuation is set to $10^{-3}$. 
In the Alfv{\'e}n wave, only $v_z$ and $B_z$ fluctuate.
The amplitude of the fluctuation of $B_z$ is set to $10^{-3}$.
The vector component parallel (perpendicular) to the wavenumber is expressed in 
terms of the subscripts $\xi$ ($\psi$) in the $(x,y)$ plane.

\begin{figure}
        \begin{center}
                \includegraphics[width=7cm]{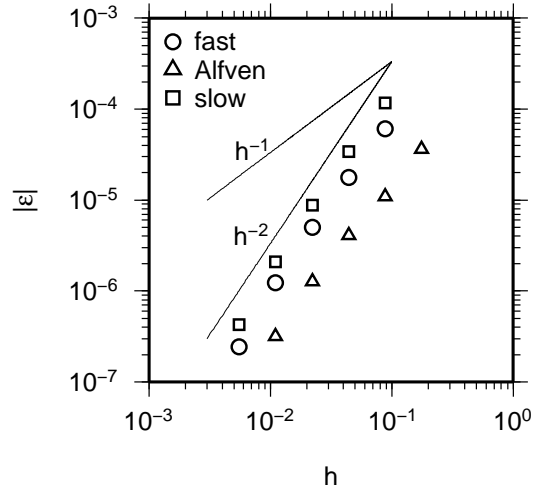}
        \end{center}
        \caption{
        Results of 
        convergence test of the fast (circles), Alfv{\'e}n (triangles), 
        and slow waves (boxes).
        The abscissa indicates the average smoothing length that represents the resolution.
        The ordinate indicates the norm of the error vector $\bmath{\epsilon}$.
        The upper and lower solid lines represent the lines of $\propto h^{-2}$ and of $\propto h^{-1}$,
        respectively.
        }
        \label{fig:convergence}
\end{figure}

As a measure of the error,
we introduce the error vector defined as 
\begin{equation}
        \bmath{\epsilon} = \frac{1}{N_\mathrm{tot}} \sum_{i=1}^{N_\mathrm{tot}} 
        | \bmath{U}_\mathrm{ref}(\bmath{x}_i) - \bmath{U}_i(\bmath{x}_i) |,
\end{equation}
where $\bmath{U}=(\rho,\;v_\xi,\;v_\psi,\;B_\psi,\;E)$ for the fast and slow waves, 
$\bmath{U}=(v_z,\;B_z)$ for the Alfv{\'e}n wave.
As a reference solution, $\bmath{U}_\mathrm{ref}$, 
we adopt the results with $N_\mathrm{tot}=512\times512$.
To eliminate the error coming from $\Delta t$, 
$\Delta t$ is set to the small value of $3\times10^{-4}$ in all resolutions. 
The error vector is evaluated after 100 time steps at various resolutions.
In Fig. \ref{fig:convergence}, the norm of the error vector is plotted as a function of 
the average smoothing length, which represents the resolution for the fast (the circles), Alfv{\'e}n 
(triangles), and slow waves (boxes).
In the scheme having second-order of spatial accuracy, $|\epsilon|$ is expected to scale as $h^2$.
Fig. \ref{fig:convergence} shows that the error is proportional to $h^2$ for all wave modes.
Therefore, it is confirmed that our GSPMHD is spatially a second-order scheme.

\subsection{Non-linear circularly polarized Alfv{\'e}n wave}\label{alf circ}
\begin{figure*}
        \begin{center}
                \includegraphics[width=12cm]{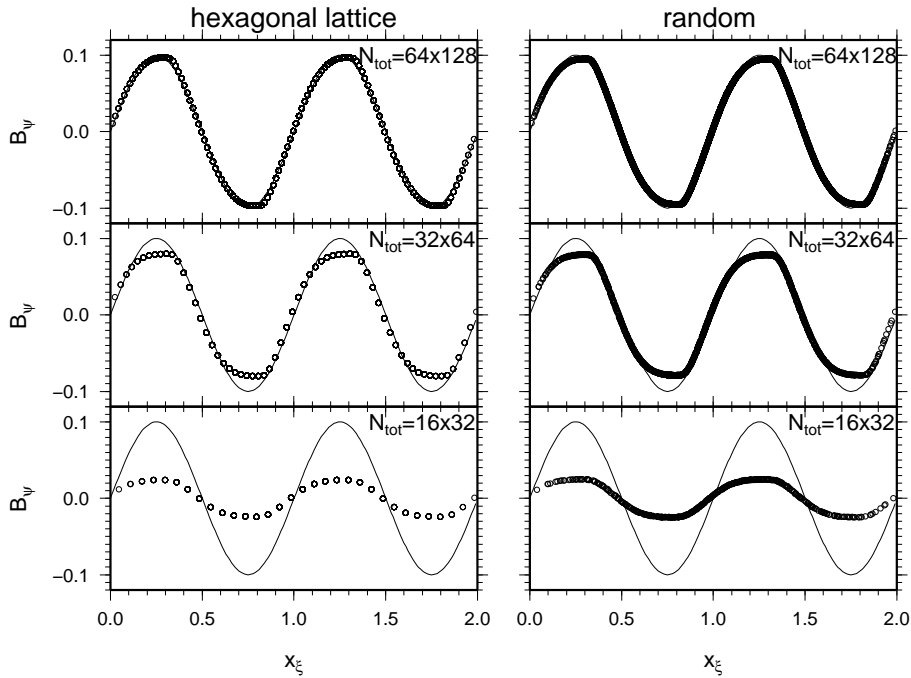}
        \end{center}
        \caption{
        Results of the circularly polarized Alfv{\'e}n wave test after 5 periods.
        The ordinates indicate the magnetic field perpendicular to the wave number, $\bmath{k}$.
        The abscissas indicate the projected coordinate in the direction of $\bmath{k}$.
        The analytic solution is shown by the solid line in each panel.
        All particles are plotted by the circles.
        The results are shown at three different resolutions $N_\mathrm{tot}=16\times32$, $32\times64$, and 
        $64\times128$ from bottom to top. The left and right panels correspond to the hexagonal lattice and th
        random distributions in the initial condition, respectively.
        }
        \label{fig:alf circ}
\end{figure*}
\begin{figure*}
        \begin{center}
                \includegraphics[width=14cm]{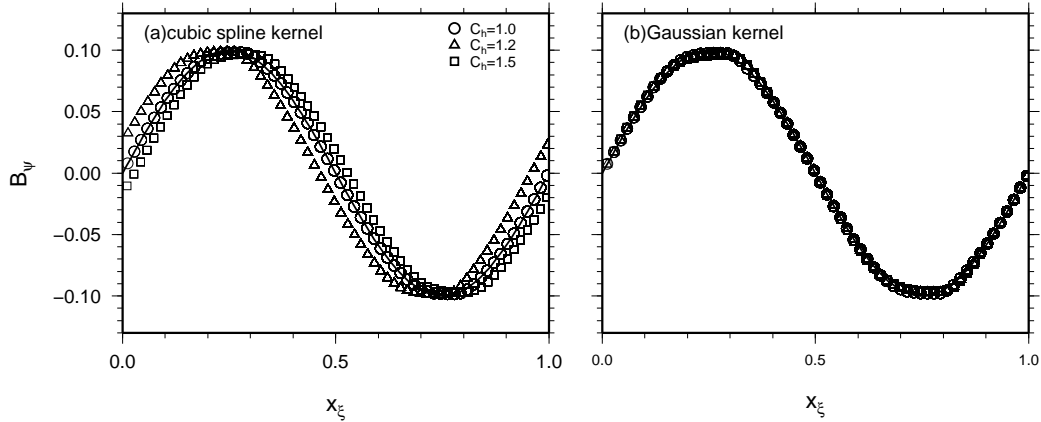}
        \end{center}
        \caption{
        Results of the circularly polarized Alfv{\'e}n wave test with (a) the cubic spline kernel and 
        (b) the Gaussian kernel after 1 period.
        The total particle number is $64\times128$. 
        The half of the calculation region $(0\le x_\xi \le 1)$ is plotted.
        The solid line incidate the analytic solution in each panel.
        In each panel, the results are shown with different smoothing length $C_h=1.0$ (the circles), 
        1.2 (the triangles), and 1.5 (the boxes) (see equation (\ref{hi})).
        }
        \label{fig:alf cubic}
\end{figure*}
\begin{figure}
        \begin{center}
                \includegraphics[width=8cm]{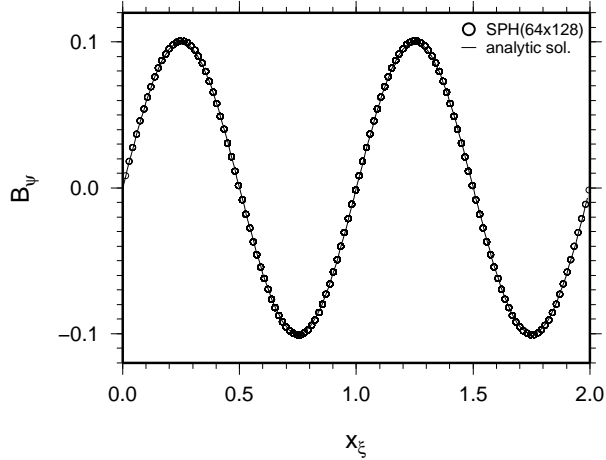}
        \end{center}
        \caption{
        Results of the circularly polarized Alfv{\'e}n wave test without a monotonicity constraint 
        for $N_\mathrm{tot}=64\times128$ (the circles) after 5 periods. The solid line indicates the analytic solution.
        }
        \label{fig:alf circ nomon}
\end{figure}
\citet{T00} investigated a non-linear circularly polarized 
Alfv{\'e}n wave that is one of the exact solutions of the non-linear MHD. 
Following \citet{T00}, we set the following initial condition.
The Alfv{\'e}n wave propagates toward an angle 
$\alpha=\pi/6$ with respect to the $x$-axis.
The initial condition is $\rho=1,\;P=0.1$, $B_\xi=1$, $v_\psi=B_\psi=0.1\sin\left( 2\pi x_\xi \right)$,
and $v_z=B_z=0.1\cos\left( 2\pi x_\xi \right)$, where $x_\xi=x\cos\alpha + y\sin\alpha$.
In order to investigate the effect of the particle distribution, 
two kind of initial particle distributions are considered. 
One is the ordered distribution (a hexagonal packed lattice), and 
the other is the random distribution that is relaxed until the density dispersion is sufficiently small. 
In each particle distribution, we calculate this test at three resolutions, $16\times32$, $32\times64$, and 
$64\times128$ particles.

The results are shown in Fig. \ref{fig:alf circ} after five periods.
The abscissa and the ordinate denote the projected coordinate $x_\xi$ 
and the perpendicular magnetic field $B_\psi$ in the $(x,y)$ plane, 
respectively. The exact solution is shown by the solid line in each panel.
All particles are plotted by the circles.
The results are shown at three different resolutions $N_\mathrm{tot}=16\times32$, $32\times64$, and 
$64\times128$ from bottom to top. The left and right panels correspond to the hexagonal lattice and the
random distributions as the initial condition, respectively.
Fig. \ref{fig:alf circ} also shows that the results with different initial particle configurations 
agree extremely well in each resolution.
\citet{PM05} performed the same test. 
In their Fig. 6, the phase error is found even in the highest resolution.
In all panels of Fig. \ref{fig:alf circ}, the phase error is not seen in our GSPMHD. 
The phase error may come from the fact that PM05 used the cubic spline kernel.
The results with the cubic spline and the Gaussian kernels after one period are shown in 
Figs. \ref{fig:alf cubic}a and \ref{fig:alf cubic}b for $N_\mathrm{tot}=64\times128$, respectively. 
To see the phase error clearly, Figs. \ref{fig:alf cubic} are plotted for $0\le x_\xi\le 1$.
In each panel, we show the results with different smoothing length $C_h=1.0$ (the circles), 
1.2 (the triangles), and 1.5 (the boxes) (see equation (\ref{hi})).
The solid line indcates the analytic solution in each panel.
From Fig. \ref{fig:alf cubic}a,  even after one period, 
one can see that the cubic spline kernel gives relatively large phase errors,
the values of which depend on the smoothing length, $C_h$.
On the other hand, the Gaussian kernel shows sufficiently small 
phase errors compared with the cubic spline kernel, and the phase errors are nearly independent of $C_h$.
Therefore, the Gaussian kernel is superior to the cubic spline kernel in the propagation of Alfv{\'e}n waves. 

Fig. \ref{fig:alf circ} shows some errors around the extremal points of $B_\psi$ 
at $x_\xi=0.25$, 0.75, 1.25, and 1.75.
This error comes from the strange ``clipped'' shape of the wave compared with the exact 
solution in Fig. \ref{fig:alf circ} while it was not found in PM05.
This is caused by the monotonicity constraint on the gradients of the physical variables, required  
for a stable description of discontinuities (see section \ref{sec:use riemann solver}).
Because the monotonicity constraint makes the scheme's spatial accuracy first-order around extremal points, 
the profile around extremal points dissipates preferentially and is flattened as seen in Fig. \ref{fig:alf circ}.
Fig. \ref{fig:alf circ nomon} shows the results without the monotonicity constraint for 
$N_\mathrm{tot}=64\times128$ (the circles) after five periods.
One can see that no deformation arises.
Deformation of the wave shape is also found in finite-volume methods using the
monotone upstream-centered scheme for conservative laws (MUSCL) method \citep{vL79}.
Apart from the wave shape, the waves in GSPMHD 
are less dissipated than those in PM05, who adopted the artificial resistivity.

\subsection{Shock Tube Problems}
MHD shock tube problems are widely used to 
test numerical codes.
In this section, we calculate two shock tube tests.

\begin{figure*}
        \begin{center}
                \includegraphics[width=17cm]{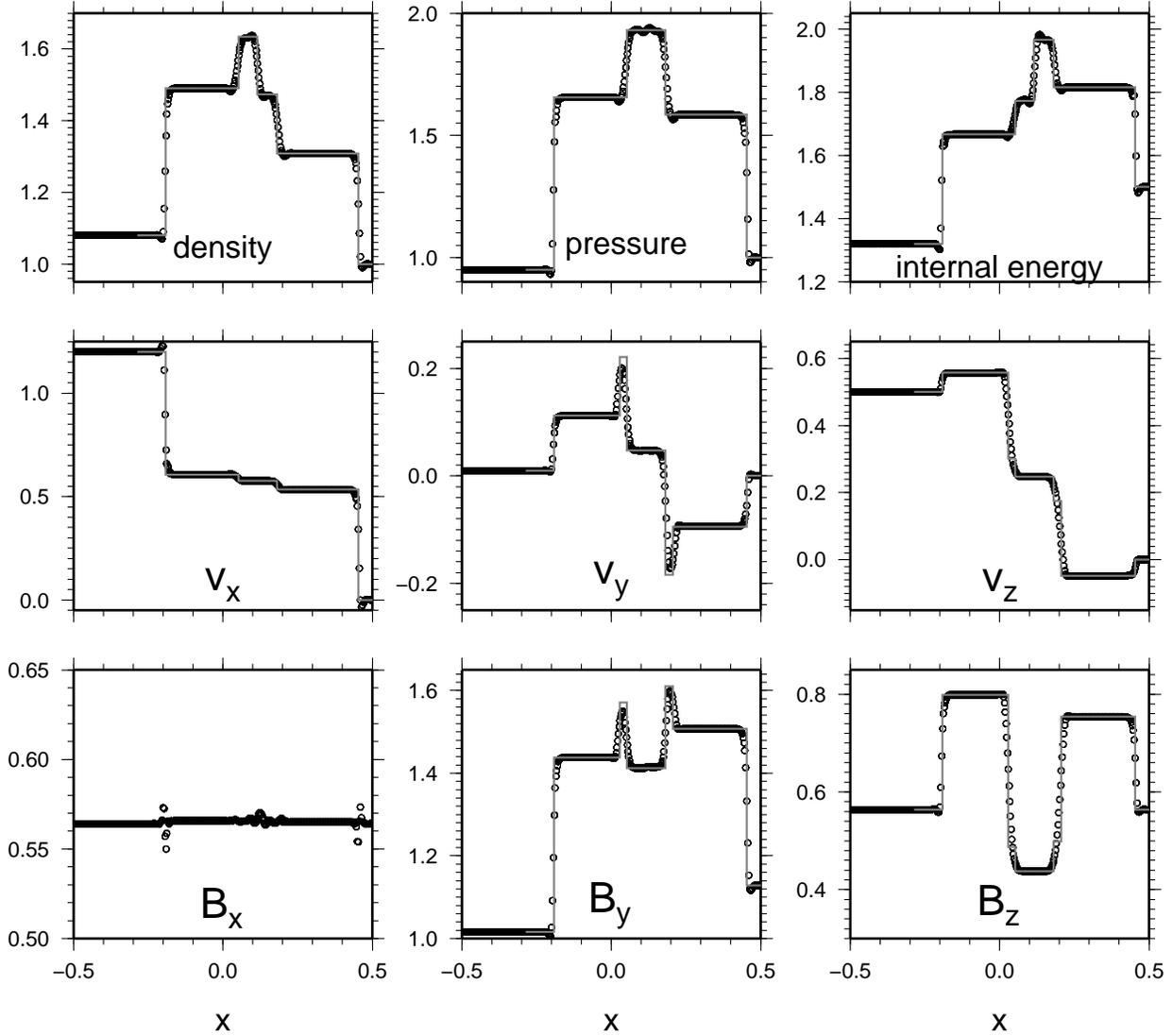}
        \end{center}
        \caption{Results of shock tube problem at $t=0.2$. The initial condition is
$(\rho,P,v_x,v_y,v_z,B_y,B_z)=($1.08, 0.95, 1.2, 0.01, 0.5, $3.6/\sqrt{4\pi}$,
$2/\sqrt{4\pi})$ for $x<0$ and $($1, 1, 0, 0, 0,$4/\sqrt{4\pi}$, $2/\sqrt{4\pi})$ 
for $x>0$ with $B_x=2/\sqrt{4\pi}$.
The circles indicate results of GSPMHD.
The solid gray lines indicate the exact solution.
        }
        \label{fig:shock3_1dim}
\end{figure*}

First, we perform a shock tube where initial states are given by 
($\rho$, $P$, $v_x$, $v_y$, $v_z$, $B_y$, $B_z$)=(1.08, 0.95, 1.2, 0.01, 0.5, $3.6/\sqrt{4\pi}$,
$2/\sqrt{4\pi})$ for $x<0$ and $($1, 1, 0, 0, 0, $4/\sqrt{4\pi}$, $2/\sqrt{4\pi})$ 
for $x>0$ with $B_x=2/\sqrt{4\pi}$.
Initially, SPH particles are distributed in a hexagonal lattice with 
the particle separation of $4\times10^{-3}$.
The particle separation in the $x$-direction for $x>0$ widens slightly 
to obtain the initial density discontinuity.
The rectangular domain is $[-0.74,0.5]\times[-3.2\times10^{-2},3.2\times10^{-2}]$.
The same shock tube problem was presented by \citet{DW94} and \citet{RJ95} in 
finite-volume methods and PM05 in SPMHD.
The exact solution consists of two fast shocks, two rotational discontinuities, 
two slow shocks, 
and one contact discontinuity.
Fig. \ref{fig:shock3_1dim} shows the results of the GSPMHD at $t=0.2$.
The solid gray lines indicate the exact solution.
One can see that GSPMHD describes all discontinuities very well.
Our GSPMHD can resolve the rotational discontinuities
and the slow shocks although they are smeared out in Fig 9 of PM05.
This may illustrate the contrast that our GSPMHD takes into account the characteristics of Alfv{\'e}n waves and 
PM05 use an artificial resistivity in the induction equation.

\begin{figure*}
        \begin{center}
                \includegraphics[width=16cm]{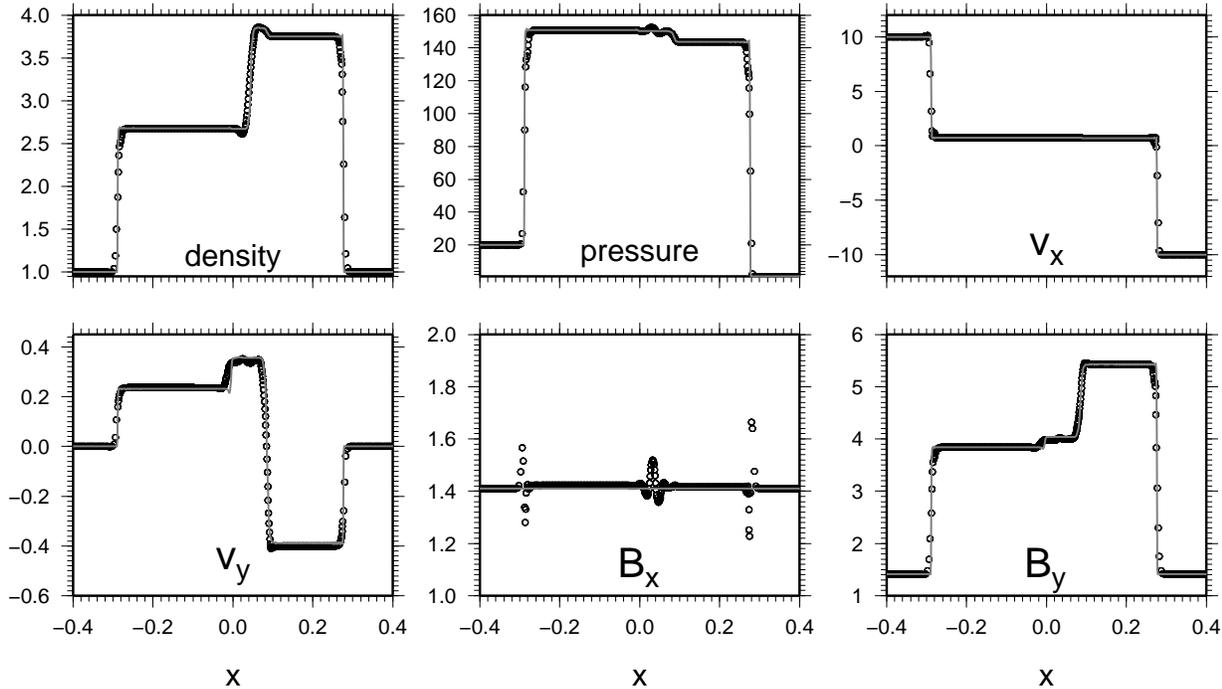}
        \end{center}
        \caption{Results of shock tube test with the initial condition 
$(\rho,P,v_x,v_y,v_z,B_y,B_z)=($1, 20, 10, 0, 0, $5/\sqrt{4\pi}$, 0) 
for $x<0$ and (1, 1, -10, 0, 0, $5/\sqrt{4\pi}$, 0) for $x>0$ 
with $B_x=5/\sqrt{4\pi}$ at $t=0.06$. The circles indicate results of GSPMHD.
The solid gray lines indicate the exact solution.
        }
        \label{fig:shock4_1dim}
\end{figure*}

Next, we perform a shock tube test contains stronger shocks than the previous one.  
The initial states are given by 
$(\rho,P,v_x,v_y,v_z,B_y,B_z)=(1,20,10,0,0,5/\sqrt{4\pi},0)$ 
for $x<0$ and $(1,1,-10,0,0,5/\sqrt{4\pi},0)$ for $x>0$ 
with $B_x=5/\sqrt{4\pi}$.
The same shock tube problem was presented by \citet{DW94}, \citet{RJ95} and \citet{T00}
in finite-volume methods.
The exact solution consists of two fast shocks, a left-propagating slow rarefaction wave, 
a right-propagating slow shock, and one contact discontinuity.
The initial particle distribution is a hexagonal lattice with 
the average particle separation of $5.4\times10^{-3}$ in $[-1,1]\times
[-7.2\times10^{-2},\;7.2\times10^{-2}]$.
Fig. \ref{fig:shock4_1dim} shows the results of the GSPMHD at $t=0.06$,
with the solid gray lines indicating the exact solution.
This figure shows that GSPMHD can reproduce the exact solution and 
describes all discontinuities better than those in PM05 and \citet{Betal06}.  
The fast shocks can be resolved by small number of particles.
In the finite-volume method, \citet{T00} reported a relatively large error of $B_x$ in his Fig. 13 where
the non-conservative method \citep{Petal99} is used.
However, even in this kind of shock tube problem with strong shocks, our scheme shows
the error in $B_x$ less than 1 percent except for the vicinity of the discontinuities.

\subsection{Orszag-Tang Vortex}
\begin{figure*}
        \begin{center}
                \includegraphics[width=14cm]{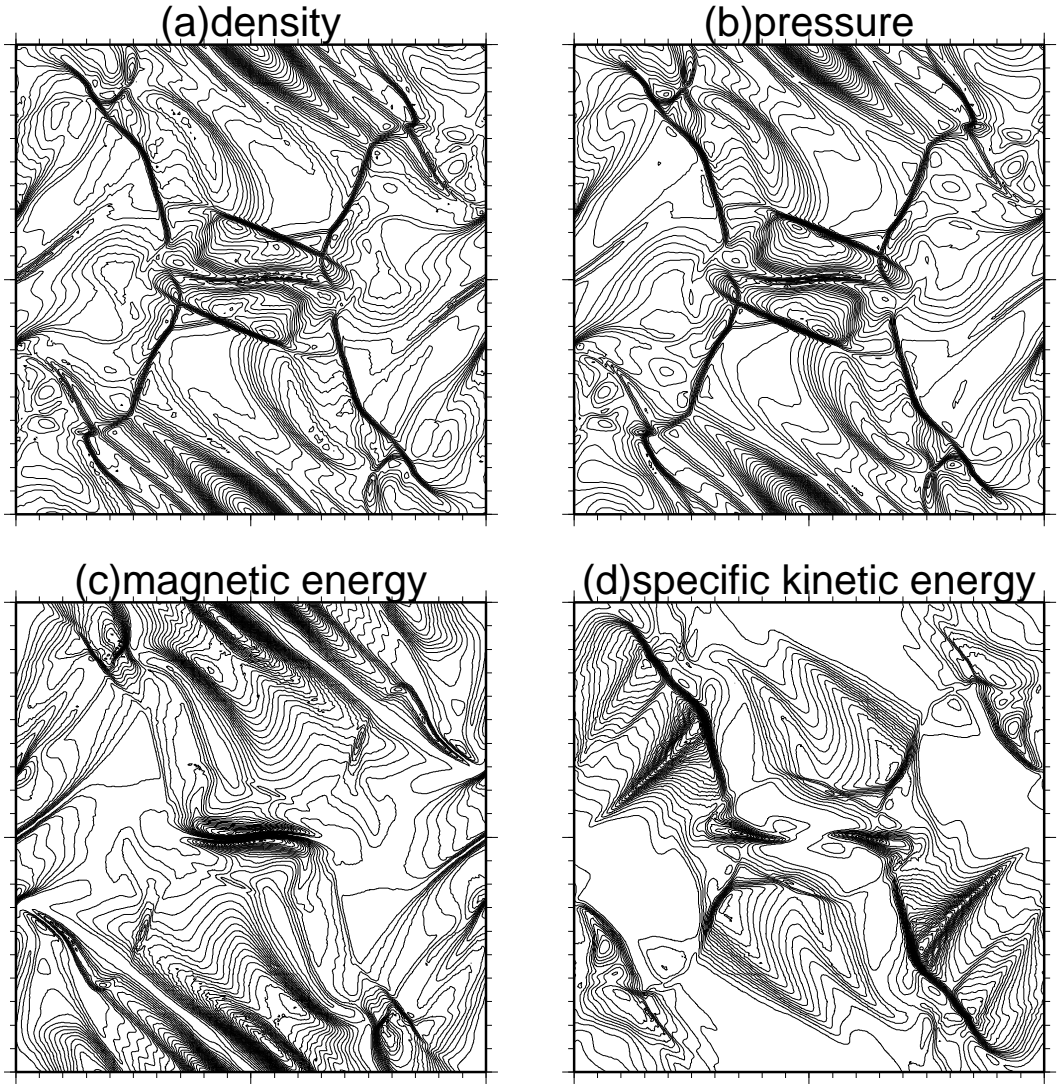}
        \end{center}
        \caption{
        Contour maps of the Orszag-Tang vortex test at $t=0.5$ for 
        (a)density, (b)pressure, (c)magnetic energy $\bmath{B}^2/2$, and (d)specific kinetic energy $\bmath{v}^2/2$.
        }
        \label{fig:otvtem}
\end{figure*}
\begin{figure*}
        \begin{center}
                \includegraphics[width=17cm]{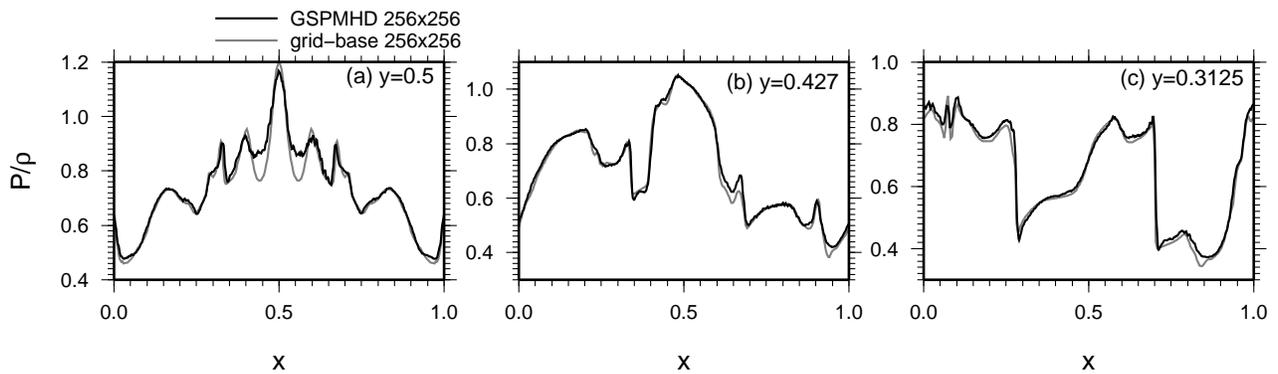}
        \end{center}
        \caption{
        (a)Horizontal slices of the temperature in Orszag-Tang vortex at $t=0.5$ taken at 
        (a)$y=0.5$, (b)$y=0.427$, and (c)$y=0.3125$. 
        The black and gray lines in each panel denote the results with the GSPMHD and 
        the finite-volume method, respectively.
        }
        \label{fig:otv1dim}
\end{figure*}
\begin{figure*}
        \begin{center}
                \includegraphics[width=15cm]{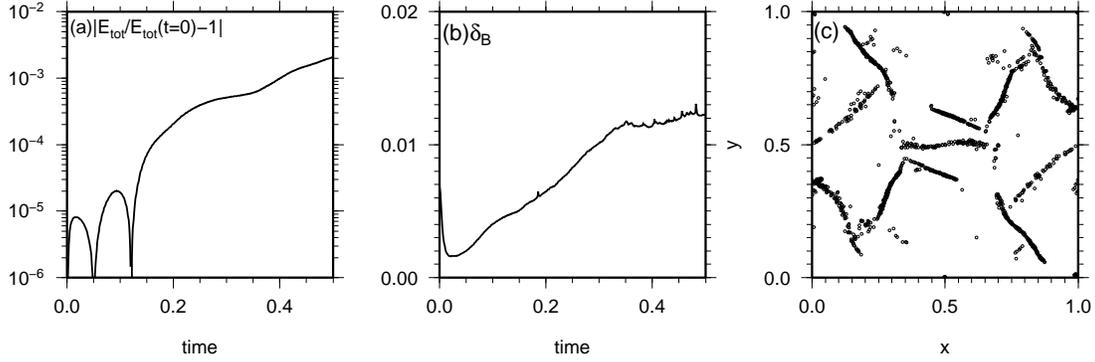}
        \end{center}
        \caption{
        Time evolution of the error of (a)the total energy $|E_\mathrm{tot}/E_\mathrm{tot}(t=0)-1|$ and 
        (b)the divergence error $\delta_\mathrm{B}$ for the Orszag-Tang Vortex test.
        (c)The spatial distribution of the SPH particles for $h_i|\bmath{\nabla}\cdot\bmath{B}|_i/|B_i|>0.05$
        }
        \label{fig:otvdivB}
\end{figure*}
The next test is the Orszag-Tang vortex problem that was originally 
investigated by \citet{OT79} in incompressible MHD flows.
This problem is a standard two-dimensional test for compressible MHD schemes \citep{T00}.
This calculation is performed in $[0,1]\times[0,1]$ domain.
In all boundaries, periodic boundary conditions are imposed.
The initial conditions are given by  $\rho = 25/(36\pi)$, $P=5/(12\pi)$,
\begin{equation}
 \bmath{v}(x,y) = \left( \sin(2\pi y),\sin(2\pi x),0 \right),\;\mathrm{and}\;\;
 \bmath{B}(x,y) = \frac{1}{\sqrt{4\pi}}\left( -\sin(2\pi y),\sin(4\pi x),0 \right).
\end{equation}
Although the initial velocity and magnetic field are not random, 
the system moves into turbulence through non-linear interaction of MHD waves.
Fig. \ref{fig:otvtem} shows that 
contour maps of the test at $t=0.5$ for 
(a)density, (b)pressure, (c)magnetic energy $\bmath{B}^2/2$, and (d)specific kinetic energy $\bmath{v}^2/2$.
Fig. \ref{fig:otvtem} can be directly compared with Fig. 22 of \citet{Setal08} and
one can see that the agreement is excellent.
To view the results quantitatively, we compare the horizontal cuts of 
the temperature for GSPMHD and a finite-volume method with HLLD Rieman solver and the constraint transport method
(provided by Dr. T. Matsumoto)
in Fig. \ref{fig:otv1dim} at (a)$y=0.5$, (b)$y=0.427$, 
and (c)$y=0.3125$. The black and grey solid lines denotes the results with the GSPMHD and 
the finite-volume method with the resolution of 256$\times$256.
One can see that the profiles between the two methods show good agreement except for 
the peak at $x=0$ in the $y=0.5$ slice.
Our scheme does not strictly conserve the total energy, $E_\mathrm{tot}=\sum_i m_i E_i$.
The time evolution of the relative error of $E_\mathrm{tot}$ is presented in 
Fig. \ref{fig:otvdivB}a.
One can see that the error of $E_\mathrm{tot}$ is sufficiently small.
Fig. \ref{fig:otvdivB}b shows 
the time evolution of the divergence error, which is maintained at an acceptable level $\sim 1$ parcent.  
The distribution of the divergence error localizes at shock fronts.
The spatial distribution of the SPH particles for $h_i|\bmath{\nabla}\cdot\bmath{B}|_i/|B_i|>0.05$
is shown in Fig. \ref{fig:otvdivB}c. 
The divergence error is highly localized at discontinuities.
This error comes from the irregular particle distribution.

\subsection{Rotor}
The MHD rotor problem was introduced by \citet{BS99} to test propagation of 
strong torsional Alfv{\'e}n waves.
The computation domain is a square unit $[-0.5,0.5]\times[-0.5,0.5]$.
This problem consists of a dense and rapidly rotating cylinder (rotor) 
embedded by a rarefied uniform medium.
The initial conditions are given by 
\begin{equation}
        \rho=10,\;\;\bmath{v}=(-2y/r_0,2x/r_0,0)\;\;\mathrm{for}\;\;
  r\equiv\sqrt{x^2+y^2}<r_0, 
\end{equation}
\begin{equation}
        \rho=1+9f(r),\;\;\bmath{v}=f(r)(-2y/r,2x/r,0)\;\;
        \mathrm{for}\;\;r_0<r<r_1,
\end{equation}
and 
\begin{equation}
        \rho=1,\;\;\bmath{v}=\bmath{0}
\end{equation}
for $r>r_1$, where $f(r)=(r_1-r)/(r_1-r_0)$, $r_0=0.1$, and $r_1=0.115$. 
The pressure and the magnetic field $P=1$, $\bmath{B}=(5/\sqrt{4\pi},0,0)$ are uniform,
and the adiabatic index is $\gamma=1.4$.
All SPH particles are assumed to have the same mass. 
Therefore, the number density of SPH particles in the rotor is larger than that in the ambient gas.
The SPH particle mass is determined so that the resolution in the ambient gas is as large as $256\times256$,
leading that the total particle number is 86968.
The initial particle distribution is constructed by using a relaxation method presented in 
\citet{Wetal95}.
\begin{figure*}
        \begin{center}
                \includegraphics[width=14cm]{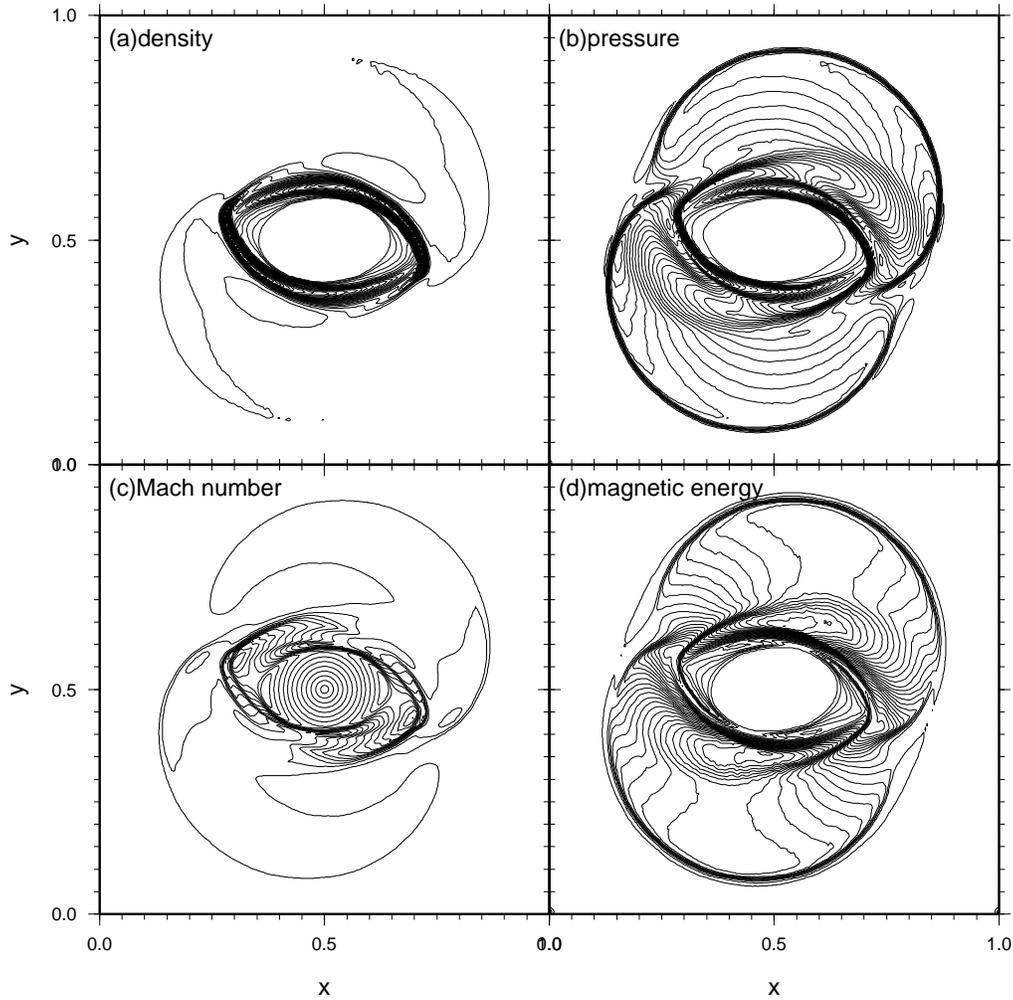}
        \end{center}
        \caption{Contour maps of rotor problem at $t=0.15$ for (a)density, (b)gas pressure,
        (c)Mach number, $|\bmath{v}|/\sqrt{\gamma P/\rho}$, and (d)magnetic energy $\bmath{B}^2/2$.
        The contour lines are the same as those in \citet{T00}.
        }
        \label{fig:rotor}
\end{figure*}
\begin{figure*}
        \begin{center}
                \includegraphics[width=16cm]{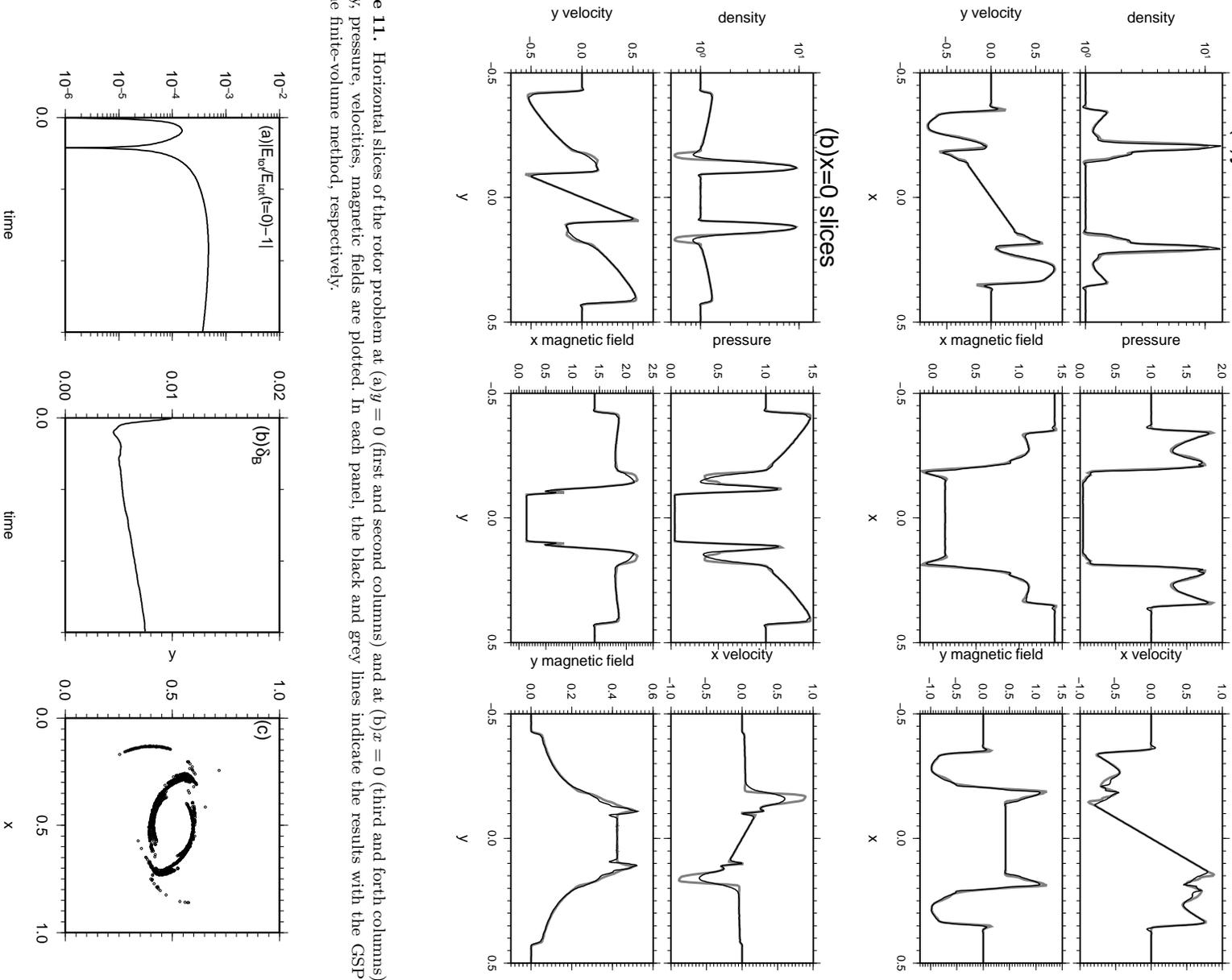}
        \end{center}
        \caption{
        Horizontal slices of the rotor problem at (a)$y=0$ (first and second columns) 
        and at (b)$x=0$ (third and forth columns).
        The density, pressure, velocities, magnetic fields are plotted.
        In each panel, the black and grey lines indicate the results with the GSPMHD and the 
        finite-volume method, respectively.
        }
        \label{fig:rotor1dim}
\end{figure*}
\begin{figure*}
        \begin{center}
                \includegraphics[width=15cm]{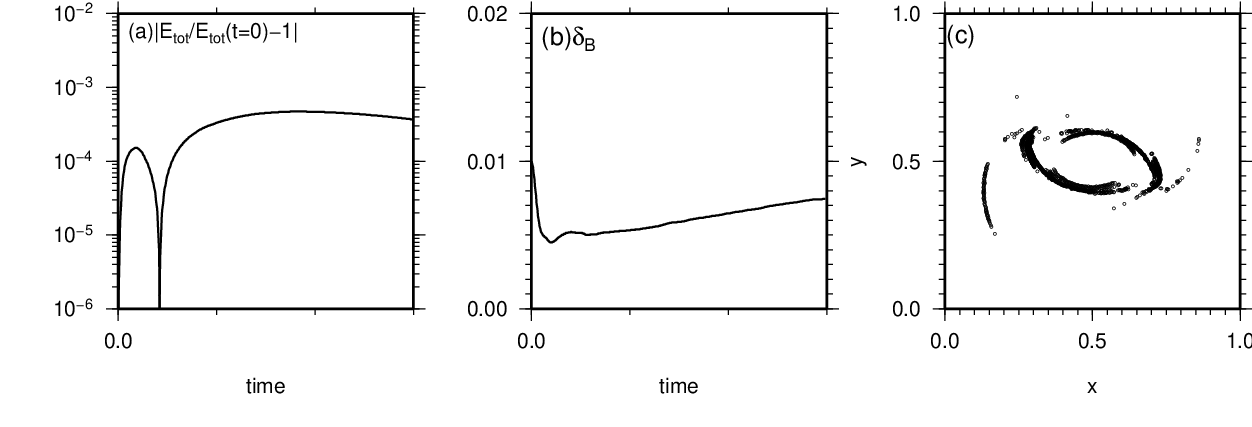}
        \end{center}
        \caption{
        The same as Fig. \ref{fig:otvdivB} but for the rotor test.
        }
        \label{fig:rotordivB}
\end{figure*}

Fig. \ref{fig:rotor} shows the contour maps of (a)density, (b)gas pressure, (c)Mach number
$|\bmath{v}|/\sqrt{\gamma P/\rho}$,
and (d)magnetic energy $\bmath{B}^2/2$ at $t=0.15$. The contour levels are the same as those in 
Fig. 18 of \citet{T00}. 
From Fig. \ref{fig:rotor}, the results agree with \citet{T00} quite well.
Compared with other SPMHD schemes, such as \citet{PM05,Betal06}, 
the contours appear to be smoother.

To compare with the finite-volume method in more detail, we plot 
horizontal slices of the rotor problem at $y=0.5$ (first row) and at $x=0.5$ (second row) in 
Fig. \ref{fig:rotor1dim}. 
In each panel, the black and grey lines indicate results with GSPMHD and the 
finite-volume method, respectively.
One can see that the agreement between the two methods is quantitatively excellent in all variables. 
Since the GSPMHD is a Lagrangian method, the resolution is better in the dense ring 
while the GSPMHD gives more diffusive results in the 
narrow region with low density ahead of the dense ring .
Fig. \ref{fig:rotordivB} is the same as Fig. \ref{fig:otvdivB} but for the rotor test.
In this test, the energy and the divergence error is maintained in a sufficiently low level. 

\subsection{Blast Wave in a Strongly Magnetized Gas}
\begin{figure*}
        \begin{center}
                \includegraphics[width=14cm]{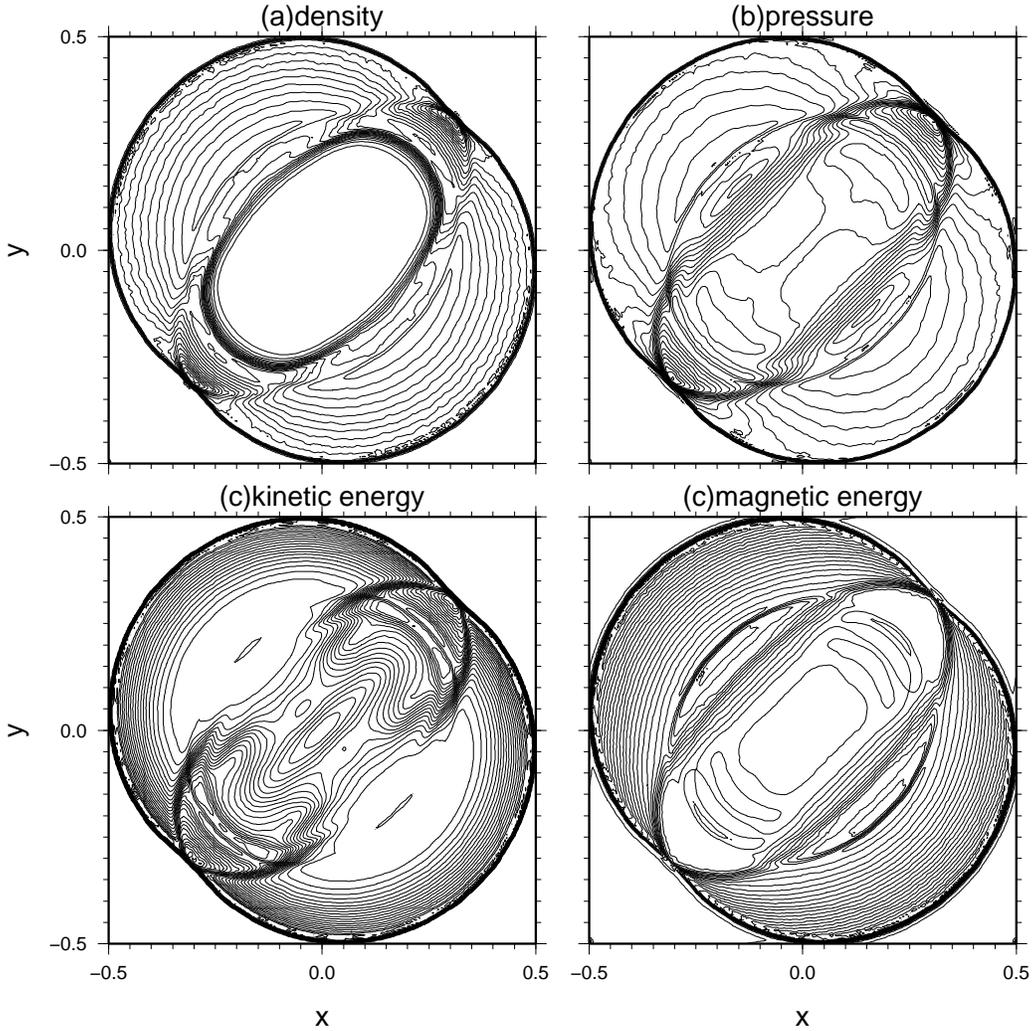}
        \end{center}
        \caption{Contour maps of the moderate $\beta$ case 
        at $t=0.15$ for (a)density, (b)gas pressure,
        (c)specific kinetic energy $(\bmath{v}^2/2)$, and (d)magnetic energy $(\bmath{B}^2/2)$.
        The 30 contour lines are shown for the ranges $0.14<\rho<2.78$, $0<P<0.95$, $0<\bmath{v}^2/2<0.37$, and 
        $0.105<\bmath{B}^2/2<1.4$.
        }
        \label{fig:blast}
\end{figure*}
\begin{figure*}
        \begin{center}
                \includegraphics[width=16cm]{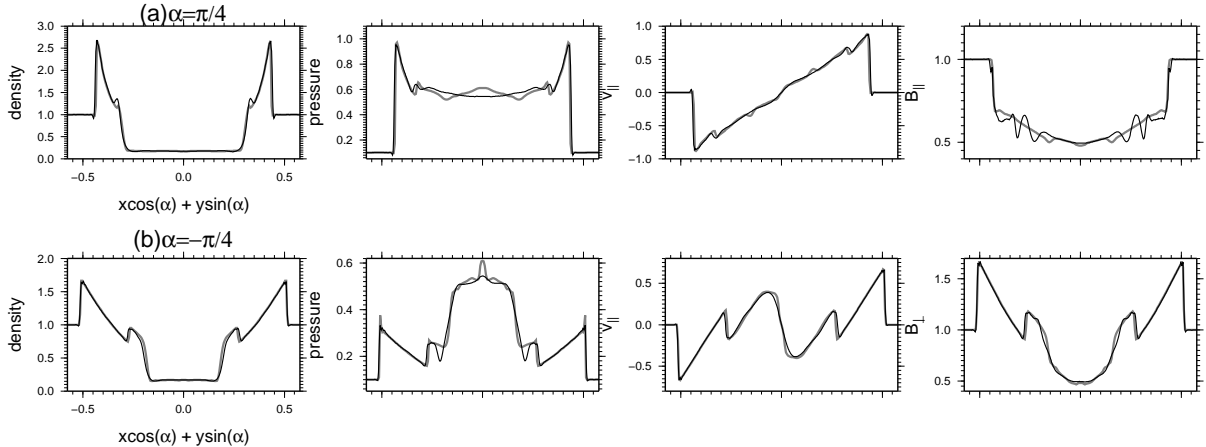}
        \end{center}
        \caption{
        Horizontal slices of the blast wave in the relatively low $\beta$ case 
        at (a)$\alpha=\pi/4$ (first column) 
        and at (b)$\alpha=-\pi/4$ (second column). 
        The density, pressure, velocities, magnetic fields are plotted.
        In each panel, the black and gray lines indicate the results with the GSPMHD and the 
        finite-volume method, respectively.
        }
        \label{fig:blast1dim}
\end{figure*}
\begin{figure*}
        \begin{center}
                \includegraphics[width=15cm]{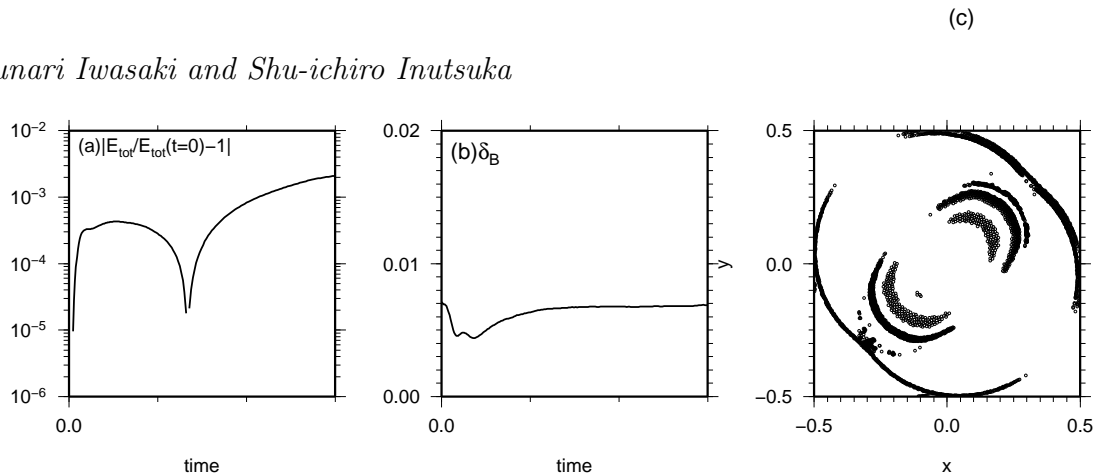}
        \end{center}
        \caption{
        The same as figure \ref{fig:otvdivB} but for the blast wave test.
        }
        \label{fig:blastdivB}
\end{figure*}

The next test is blast waves that propagates into a strongly 
magnetized gas \citep{BS99,LD00}.
The calculation region is a square domain of $[-0.5,0.5]\times[-0.5,0.5]$.
In this problem, an overpressured hot region with pressure of $P_\mathrm{hot}$ 
is set within $r<r_0$, where $r=\sqrt{x^2+y^2}$.
Around the hot central region there is a rarefied ambient 
gas with a pressure of $P_\mathrm{amb}$. The density $\rho=1$ is spatially uniform. 
The initial uniform magnetic field $\bmath{B}=B_0(1/\sqrt{2},1/\sqrt{2},0)$ 
makes the angle $\pi/4$ with the $x$-axis.
Here, $r_0$, $P_\mathrm{hot}$, $P_\mathrm{amb}$, and $B_0$ are parameters.
We consider two cases: moderate-$\beta$ case and low-$\beta$ case.

\subsubsection{Moderate-$\beta$ Case}
We adopts $r_0=0.1$, $P_\mathrm{hot}=10$, $P_\mathrm{amb}=0.1$, and $B_0=1$.
Therefore, the plasma $\beta$ of the ambient gas is as low as 0.2.
These parameters were adopted in \citet{GS05}.
Fig. \ref{fig:blast} shows that the contour maps of the blast wave at $t=0.15$ for (a)$\rho$,
(b)$P$, (c)$\bmath{v}^2/2$, and (d)$\bmath{B}^2/2$.
The total particle number is as large as $256\times256$.
One can see the shock structures around the elongated hot bubble along $\bmath{B}_0$.
The fast (slow) shock propagates toward the direction parallel (perpendicular) to $\bmath{B}_0$. 
This figure can be directly compared with the bottom column of Fig. 28 in \citet{Setal08}.
One can see that the contour maps is quite similar to those in \citet{Setal08} except for
the central rarefied hot bubble where GSPMHD is more diffusive owing to its Lagrangian nature.

For comparison with the finite-volume method in detail, we consider slices of physical 
variables passing through the centre $(0,0)$.
To characterize the direction of the slice, 
we introduce an angle $\alpha$ that is the angle between the slice and the $x$-axis.
Fig. \ref{fig:blast1dim} shows the results for the cases with $\alpha=\pi/4$ and $\alpha=-\pi/4$,
which correspond to the directions parallel and perpendicular to the initial magnetic field, respectively.
The subscripts $\parallel$ and $\perp$ represent the components 
parallel and perpendicular to the direction of the slice.
The black and grey lines indicate the results with GSPMHD and the finite-volume method, respectively.
One can see that the results with GSPMHD agree with those of the finite-volume method 
very well except in the central region, as mentioned above.
In the profile of the parallel magnetic field $B_\parallel$ for $\alpha=\pi/4$,
there are some wiggles near the contact discontinuity.
This comes from the pressure jump in the initial condition, and 
does not serious because $B_\parallel$ agrees with that in the finite-volume method in the other places.
If the pressure jump is smoother, the wiggle becomes small.
Fig. \ref{fig:blastdivB} is the same as Fig. \ref{fig:otvdivB} but for the blast wave test.
In this test, the energy and the divergence error is maintained in a sufficiently low level. 

\subsubsection{Low $\beta$ Case}
\begin{figure*}
        \begin{center}
                \includegraphics[width=12cm]{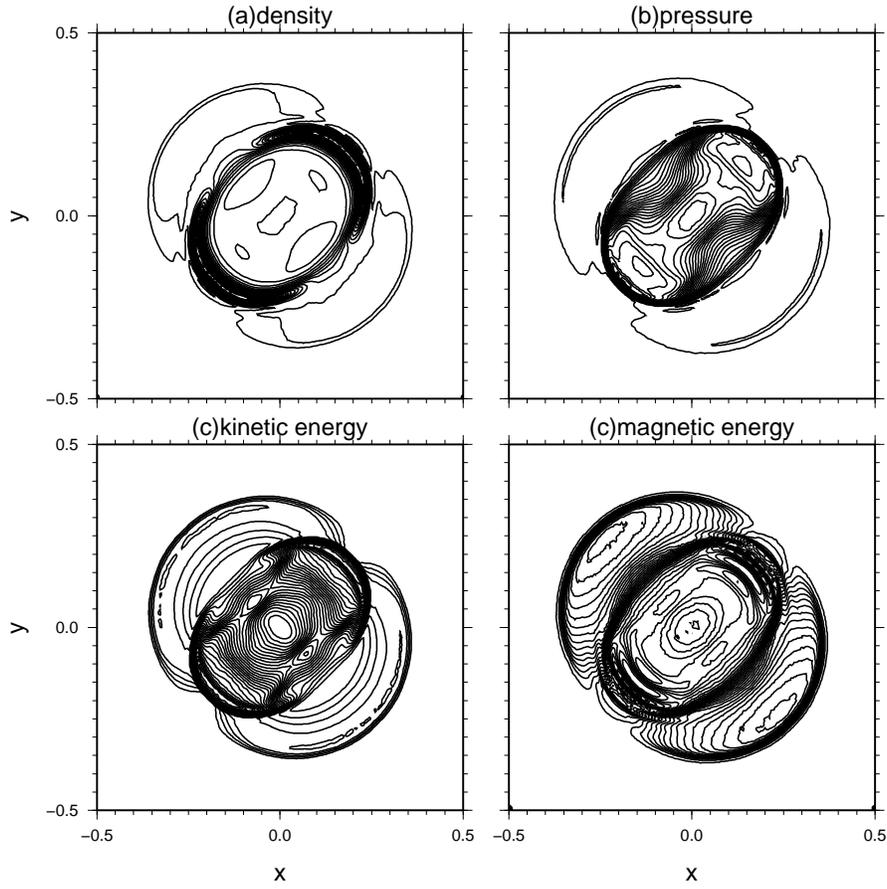}
        \end{center}
        \caption{
        The same as Fig. \ref{fig:blast} but for the low $\beta$ case at $t=0.02$. 
        The 30 contour lines are shown for the ranges $0.233<\rho<3.31$, $32.1<P<1.1$, $0<\bmath{v}^2/2<13$, and 
        $24.5<\bmath{B}^2/2<76$.
        }
        \label{fig:blast_strong}
\end{figure*}
\begin{figure*}
        \begin{center}
                \includegraphics[width=15cm]{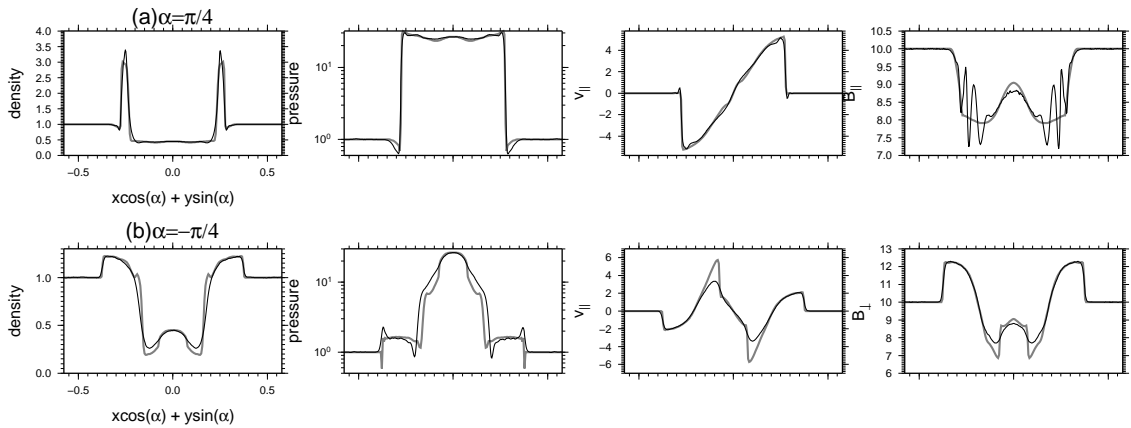}
        \end{center}
        \caption{
        The same as Fig. \ref{fig:blast1dim} but for the low $\beta$ case at $t=0.02$.
        }
        \label{fig:blast1dim_strong}
\end{figure*}
We adopts $r_0=0.125$, $P_\mathrm{hot}=100$, $P_\mathrm{amb}=1$, and $B_0=10$.
Therefore, the plasma $\beta$ of the ambient gas is as low as 0.02.
These parameters were adopted in \citet{LD00} and \citet{GS05}.
Fig. \ref{fig:blast_strong} is the same as Fig. \ref{fig:blast} but for the low $\beta$ case.
One can see a stronger slow shock along the initial magnetic field than that in the previous case.
Fig. \ref{fig:blast1dim_strong} shows the slices of the physical variables along $\alpha=\pm\pi/4$ 
with respect to the $x$-axis, and  
is the same as Fig. \ref{fig:blast1dim} but for the low $\beta$ case.
One can see that the results of GSPMHD coincide with those of the finite-volume method very well 
also in the low $\beta$ case.

\section{Summary}\label{sec:summary}
In this paper, we developed a new SPMHD scheme with the Godunov method. 
To take into account the physical dissipation, 
we consider the non-linear RP with magnetic pressure and the MOC in the 
interaction between the SPH particles instead of artificial dissipation used in previous works.
Using the MUSCL method, the spatial accuracy of our scheme attains second order accuracy $O(h^2)$ that 
is confirmed in the convergence test of linear MHD waves (see section \ref{sec:test}).
From several test calculations in section \ref{sec:test}, 
it is confirmed that our method can capture all MHD discontinuities more accurately 
than previous proposed methods.
The GSPMHD can provide results comparable to finite-volume methods with approximate Riemann solvers.
We will apply the GSPMHD to astrophysical problems where the Lagrangian description has advantages.

The GSPMHD described in this paper loses strict conservation property with respect to both momentum and energy.
This is to avoid the tensile instability in strong magnetic fields.
Although conservation errors are 
sufficiently small in the test calculations (see section \ref{sec:test}), the non-conservative formulation 
can be problematic in long-term calculations. Thus, 
further investigations are needed to improve the conservation property together with the better performance 
in low-$\beta$ plasma cases.

In the GSPMHD, the Gaussian kernel is used. Other kernel functions (e.g., the cubic spline kernel)
can be applied easily in our GSPMHD if one use equation (\ref{Fij monaghan}) 
or other symmetrization of the kernel function.
However, in the SPMHD, choice of kernel functions may be important in contrast to the HD case.
In section \ref{alf circ}, it is shown that the cubic spline kernel brings relatively large phase errors 
into the propagation of Alfv{\'e}n waves.
In shock tube tests, we confirm that the results with the cubic spline kernel are 
worse compared with the Gaussian kernel.
To obtain reasonable results with the cubic spline kernel, one need a large neighbours 
$h_i>1.5(m_i/\rho_i)^{1/2}$ in the two-dimensional code as suggested in PM05. 
Thus, we recommend the Gaussian kernel in the GSPMHD.

\section*{Acknowledgments}
We thank the referee, Dr. Daniel Price for many constructive comments 
that improved the paper. We thank Dr. Takuma Matsumoto for variable discussions and 
providing his results of test calculations using 
the HLLD$+$CT method. We also thank Dr. Takeru K. Suzuki and Dr. Toru Tsuribe for many constructive discussions.
This work was supported by Grants-in-Aid for Scientific Research
from the MEXT of Japan (K.I.:22864006; S.I.:18540238 and 16077202). 


\appendix
\section{Riemann Solver with Magnetic Pressure}\label{riemann}
\begin{figure}
        \begin{center}
                \includegraphics[width=8cm]{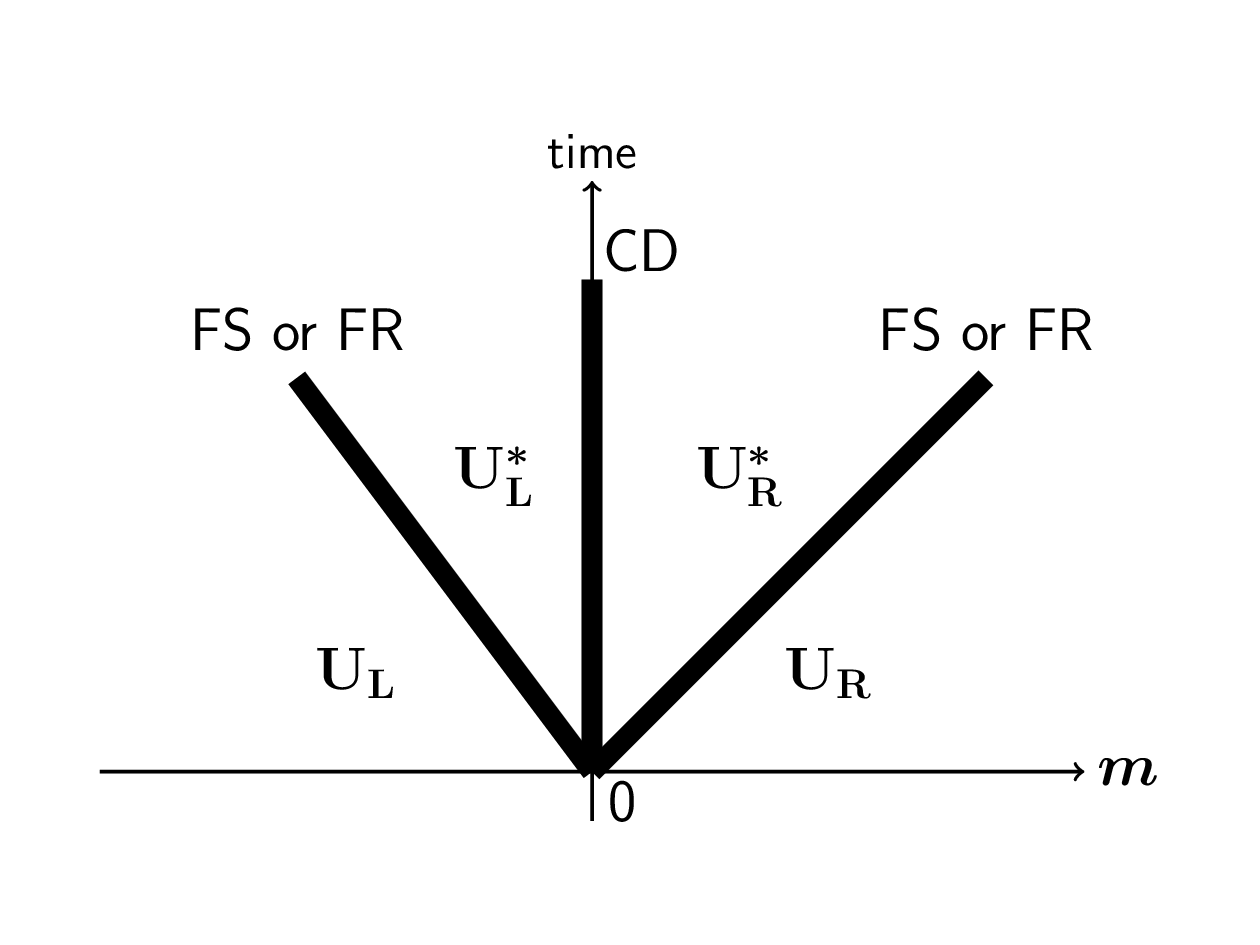}
        \end{center}
        \caption{Schematic picture of the Riemann problem.
        ``CD'', ``FS'', and ``FR'' denote the contact discontinuity, the fast shock, and 
        the fast rarefaction wave.
        }
        \label{fig:riemann}
\end{figure}

In this Appendix, we present the non-linear Riemann solver.
Fig. \ref{fig:riemann} shows that the schematic picture of the non-linear RP.
Initially, we consider two uniform states $\bmath{U}_\mathrm{L},\;\bmath{U}_\mathrm{R}$ 
that are separated by the discontinuity at $m=0$, where 
$\bmath{U} = (\rho,\;P,\;\bmath{v},\;\bmath{B}_\perp)$ and $m\equiv \int_0^s \rho ds$ 
is the mass coordinate. 
The magnetic field in the $s$-direction, $B_\parallel$,
is assumed to be zero.
The RP depends only on $\bmath{B}_\perp^2$.
Since $\partial \bmath{v}_\perp/\partial t=0$ for $B_\parallel=0$,
$\bmath{v}_\perp$ is constant spatially and temporary in each side even if $\bmath{v}_\perp$ has 
a discontinuity at $m=0$. 
Therefore, $\bmath{v}_\perp$ does not affect the RP, suggesting that we can set $\bmath{v}_\perp=0$ 
without loss of generality.

In this RP, the fast shock (FS) or the fast rarefaction (FR) waves propagate outward as shown in 
Fig. \ref{fig:riemann}. This configuration is the same as the RP
used in the Godunov method of the HD \citep{G59}.
The RP is separated into the left and the right intermediate 
state, $\bmath{U}_\mathrm{L}^*$ and $\bmath{U}_\mathrm{R}^*$,
by the contact discontinuity (CD). At the CD, since the total pressure and the velocity is 
continuous, one can get the following relations,
\begin{equation}
        P_\mathrm{t}^* \equiv P_\mathrm{L}^* + \frac{\left(B_\mathrm{\perp L}^*\right)^2}{2}
                            = P_\mathrm{R}^* + \frac{\left(B_\mathrm{\perp R}^*\right)^2}{2}
\end{equation}
\begin{equation}
        v_\parallel^* \equiv v_\mathrm{\parallel L}^* = v_\mathrm{\parallel R}^* 
\end{equation}
In order to take into account the FR exactly, we need the numerical 
integration that is computationally expensive.
Therefore, we treat the FR as ``rarefaction shock''.
This mean that the shock jump condition is used in the case of 
$P_\mathrm{t}^*<P_\mathrm{t L}$ or $P_\mathrm{t R}$.
This treatment is reasonably accurate because 
the tangential lines of the Hugoniot curve and 
the adiabatic curve coincide at any point in $(\rho,\;P_\mathrm{t})$ plane.

From equation (\ref{basic eq}), the relations between the jump of the total pressure and velocity 
across the right-facing fast shock and the left-facing fast shock 
are given by 
\begin{equation}
  P_\mathrm{t}^* - P_\mathrm{tR} = M_\mathrm{R}\left( v_\parallel^* - v_\mathrm{\parallel R} \right),\;\;
  \mathrm{and}\;\;
  \label{right}
  P_\mathrm{t}^* - P_\mathrm{tL} = - M_\mathrm{L}\left( v_\parallel^* - v_\mathrm{\parallel L} \right),
\end{equation}
respectively,
where $M_\mathrm{R}$ and $M_\mathrm{L}$ are the Lagrangian speeds of the 
right-facing and the left-facing shocks, respectively.
From equation (\ref{right}), the total pressure and the velocity in the intermediate 
state are given by 
\begin{equation}
        P_\mathrm{t}^* 
        = \frac{1}{1/M_\mathrm{R} + 1/M_\mathrm{L}}
        \left[ \frac{P_\mathrm{t R}}{M_\mathrm{R}} + \frac{P_\mathrm{t L}}{M_\mathrm{L}}
        - \left( v_\mathrm{\parallel R} - v_\mathrm{\parallel L} \right)\right],
        \label{riemann P}
\end{equation}
\begin{equation}
        v_\parallel^* 
        = \frac{1}{M_\mathrm{R} + M_\mathrm{L}}
        \left[ M_\mathrm{R} v_\mathrm{\parallel R} + M_\mathrm{L} v_\mathrm{\parallel L} 
        - \left( P_\mathrm{t R} - P_\mathrm{t L} \right)\right].
        \label{riemann v}
\end{equation}
The Lagrangian shock speeds $M_\mathrm{L}$, $M_\mathrm{R}$ are 
derived from the jump conditions across the shock,
\begin{equation}
         \left[ \rho v_\parallel \right] = 0,\;\;
         \left[ \rho v_\parallel^2  +P+\frac{B_\perp^2}{2} \right] = 0,
        \label{jump1}
\end{equation}
\begin{equation}
        \left[ v_\parallel B_\perp \right]=0,\;\;
        \left[ \left(\frac{\rho v_\parallel^2}{2} + \frac{\gamma P}{\gamma-1}\right)
        v_\parallel
        + v_\parallel B_\perp^2 \right]=0.
        \label{jump2}
\end{equation}
Using equations (\ref{jump1}) and (\ref{jump2}), one can get 
\begin{equation}
        M_a^2 = \frac{\rho_a}{4}\left[ 
        \left( \gamma-3 \right)P_\mathrm{ta} + \left( \gamma+3 \right) P_t^*
        - \left( \gamma-2 \right)B_\mathrm{\perp a}^2 + \sqrt{D_a}
        \right],
        \label{lag shock}
\end{equation}
\begin{equation}
        D_a = \left\{ \left( \gamma+1 \right)P_\mathrm{ta} + \left( \gamma-1 \right)P_\mathrm{t}^* \right\}^2
        - \left( \gamma-2 \right)B_\mathrm{\perp a}^2 \left\{ 2\left( \gamma-3 \right)P_\mathrm{ta}
        + 2(\gamma+3)P_\mathrm{t}^* - \left( \gamma-2 \right)B_\mathrm{\perp a}^2\right\},
\end{equation}
where $a=$L and R.
From equation (\ref{lag shock}), since $M_\mathrm{L}$ and $M_\mathrm{R}$
depend on $P_\mathrm{t}^*$,
equation (\ref{riemann P}) is non-linear with respect to $P_\mathrm{t}^*$.
Therefore, we solve equation (\ref{riemann P}) iteratively by the following procedure.
First, the total pressure $P_\mathrm{t}^{*(1)}=(P_\mathrm{t R}+P_\mathrm{t L})/2$ is inserted into 
the right hand side of equation (\ref{riemann P}). 
Then, we can get $P_\mathrm{t}^{*(2)}$ that is also inserted into equation (\ref{riemann P}) to 
get $P_\mathrm{t}^{*(3)}$.
The iteration is continued until the desired accuracy is reached.
Finally, the velocity $v_\parallel^*$ is obtained from equation (\ref{riemann v}).

\section{Method of Characteristics}\label{moc}
\begin{figure}
        \begin{center}
                \includegraphics[width=8cm]{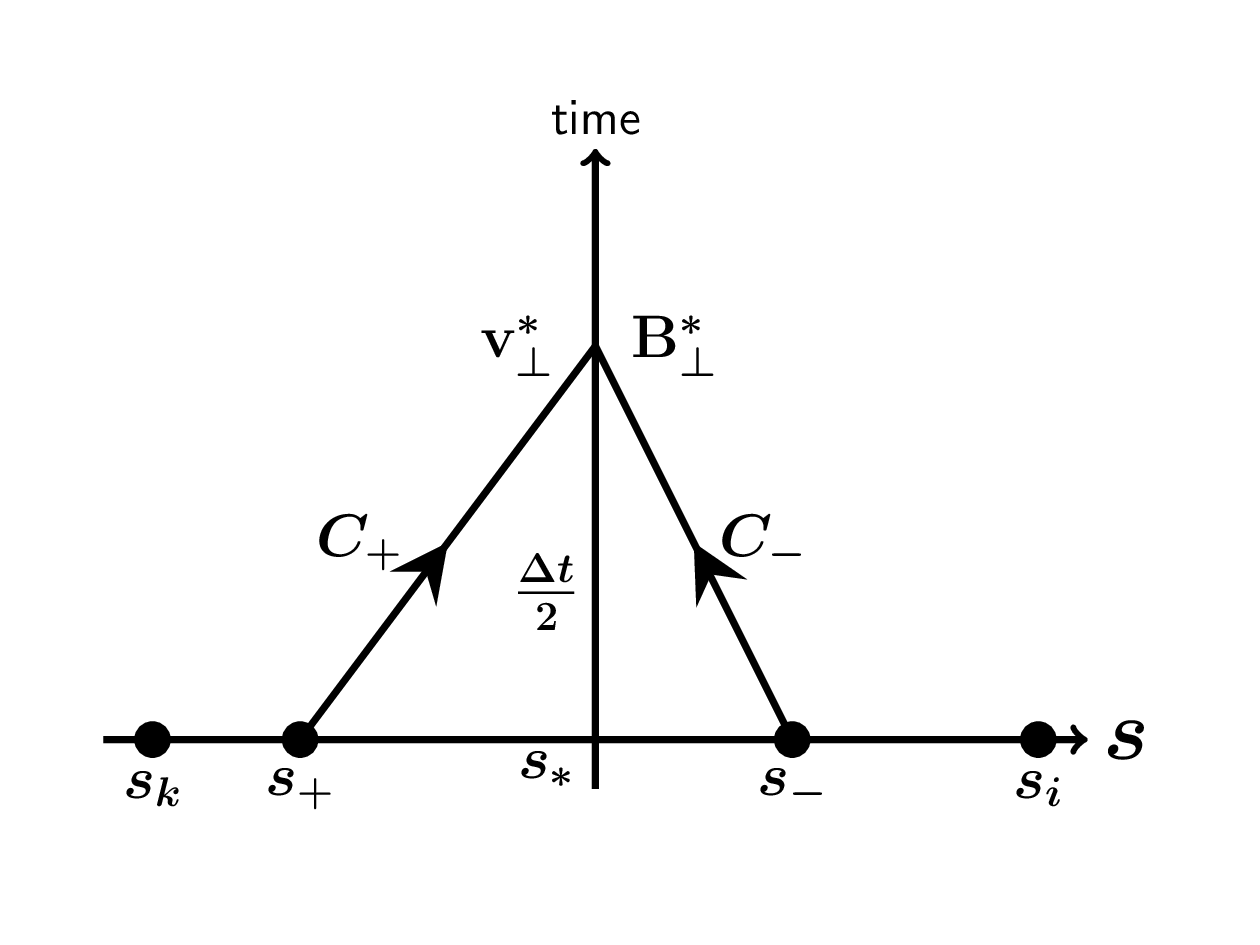}
        \end{center}
        \caption{Schematic picture of the method of characteristics for $B_\parallel>0$}
        \label{fig:moc}
\end{figure}
In this Appendix, MOC is briefly reviewed. 
We consider the propagation of the Alfv{\'e}n wave along the $s$-axis.
For simplicity, we consider the case with $B_\parallel>0$.
More general expression including the case with $B_\parallel<0$ is presented later.
The MHD equations for a one-dimensional (the $s$-direction) incompressible fluid are given by 
\begin{equation}
    \frac{d\bmath{v}_\perp}{dt} = \frac{B_\parallel}{\rho} 
    \frac{\partial \bmath{B}_\perp}{\partial s}\;\;\mathrm{and}\;\;
        \frac{d\bmath{B}_\perp}{dt} = B_\parallel \frac{\partial \bmath{v}_\perp}{\partial s}.
    \label{moc eq1}
\end{equation}
equations (\ref{moc eq1}) can be written as 
\begin{equation}
        \left( \frac{d\bmath{v}_\perp}{dt} - \frac{1}{\sqrt{\rho}}
        \frac{d\bmath{B}_\perp}{dt}\right) + 
        \frac{B_\parallel}{\sqrt{\rho}}\left( \frac{\partial \bmath{v}_\perp}{\partial s} 
        - \frac{1}{\sqrt{\rho}}\frac{\partial \bmath{B}_\perp}{\partial s}\right) = 0,
     \label{cp}
\end{equation}
and
\begin{equation}
        \left( \frac{d\bmath{v}_\perp}{dt} + \frac{1}{\sqrt{\rho}}
        \frac{d\bmath{B}_\perp}{dt}\right) - 
        \frac{B_\parallel}{\sqrt{\rho}}\left( \frac{\partial \bmath{v}_\perp}{\partial s} 
        + \frac{1}{\sqrt{\rho}}\frac{\partial \bmath{B}_\perp}{\partial s}\right) = 0.
     \label{cm}
\end{equation}
From equations (\ref{cp}) and (\ref{cm}), one can see that 
\begin{equation}
        d \bmath{J}_+= d \bmath{v}_\perp - \frac{d\bmath{B}_\perp}{\sqrt{\rho}},\;\;\mathrm{and}\;\;
        d \bmath{J}_-= d \bmath{v}_\perp + \frac{d\bmath{B}_\perp}{\sqrt{\rho}}
        \label{J+-}
\end{equation}
are constant on a trajectories with $ds/dt=B_\parallel/\sqrt{\rho}$
and $ds/dt=-B_\parallel/\sqrt{\rho}$, respectively.
Fig. \ref{fig:moc} shows the schematic picture.
The partially updated values $\bmath{v}_\perp^{*}$ 
and $\bmath{B}_\perp^{*}$ at $t+\Delta t/2$ can be obtained 
by extrapolate back in time along $C_+$ and $C_-$ to the present time step where 
all variables are known. The positions $s_+$ and $s_-$ are the foot 
points of $C_+$ and $C_-$ intersect 
at $s_*$ on $t+\Delta t/2$, respectively (see Fig. \ref{fig:moc}).
Using the Riemann invariant $d\bmath{J}_\pm$, the characteristic equations along $C_+$ and $C_-$ are given by 
\begin{equation}
        \bmath{v}_{\perp}^{*} - \bmath{v}_\perp^+ - \frac{\bmath{B}_{\perp}^{*} 
        - \bmath{B}_\perp^+} 
        {\sqrt{\rho^+}}= 0,\;\;\mathrm{and}\;\;
        \bmath{v}_{\perp}^{*} 
        - \bmath{v}_\perp^- + \frac{\bmath{B}_{\perp}^{*} - \bmath{B}_\perp^-} 
        {\sqrt{\rho^-}}= 0,
        \label{eq1}
\end{equation}
respectively.
From equations (\ref{eq1}), $\bmath{B}_\mathrm{\perp}^*$ and 
$\bmath{v}_\mathrm{\perp}^*$  are given by
\begin{equation}
        \bmath{B}_{\perp}^{*} = 
        \left( \frac{1}{\sqrt{\rho^+}} + \frac{1}{\sqrt{\rho^-}} \right)^{-1}\left[ 
        \frac{\bmath{B}_{\perp}^+}{\sqrt{\rho^+}}  + \frac{\bmath{B}_{\perp}^-}{\sqrt{\rho^-}} 
        - \bmath{v}_\perp^+ + \bmath{v}_\perp^- \right]
\end{equation}
and 
\begin{equation}
        \bmath{v}_{\perp}^{*} = 
        \frac{ \sqrt{\rho^+}\bmath{v}_{\perp}^+  + \sqrt{\rho^-}\bmath{v}_{\perp}^-
        - \bmath{B}_\perp^+ + \bmath{B}_\perp^-  }{\sqrt{\rho^+} + \sqrt{\rho^-} },
\end{equation}
respectively.
So far, we consider only for the case with $B_\parallel>0$.
For the case with $B_\parallel<0$, the Alfv{\'e}n wave propagates 
in the opposite direction of the $s$-axis. 
Therefore, the positions of $s^+$ and $s^-$ replace each other.
The general expressions are given by 
\begin{equation}
        \bmath{B}_{\perp}^{*} = 
        \left( \frac{1}{\sqrt{\rho^+}} + \frac{1}{\sqrt{\rho^-}} \right)^{-1}\left[ 
        \frac{\bmath{B}_{\perp}^+}{\sqrt{\rho^+}}  + \frac{\bmath{B}_{\perp}^-}{\sqrt{\rho^-}} 
        + \mathrm{sgn}(B_\parallel)\left(-\bmath{v}_\perp^+ + \bmath{v}_\perp^- \right)\right]
        \label{moc mag}
\end{equation}
and 
\begin{equation}
        \bmath{v}_{\perp}^{*} = 
        \frac{ \sqrt{\rho^+}\bmath{v}_{\perp}^+  + \sqrt{\rho^-}\bmath{v}_{\perp}^-
        + \mathrm{sgn}(B_\parallel)\left(-\bmath{B}_\perp^+ + \bmath{B}_\perp^-\right)  }
        {\sqrt{\rho^+} + \sqrt{\rho^-}},
        \label{moc vel}
\end{equation}
where $\mathrm{sgn}(B_\parallel)$ is the sign of $B_\parallel$.
In actual calculations, as $B_\parallel$, we use the following simple average value, 
$\left( B_{\parallel,i} + B_{\parallel,j}\right)/2$.
\section{Monotonicity Constraint}\label{app:mono}
\begin{figure}
        \begin{center}
                \includegraphics[width=8cm]{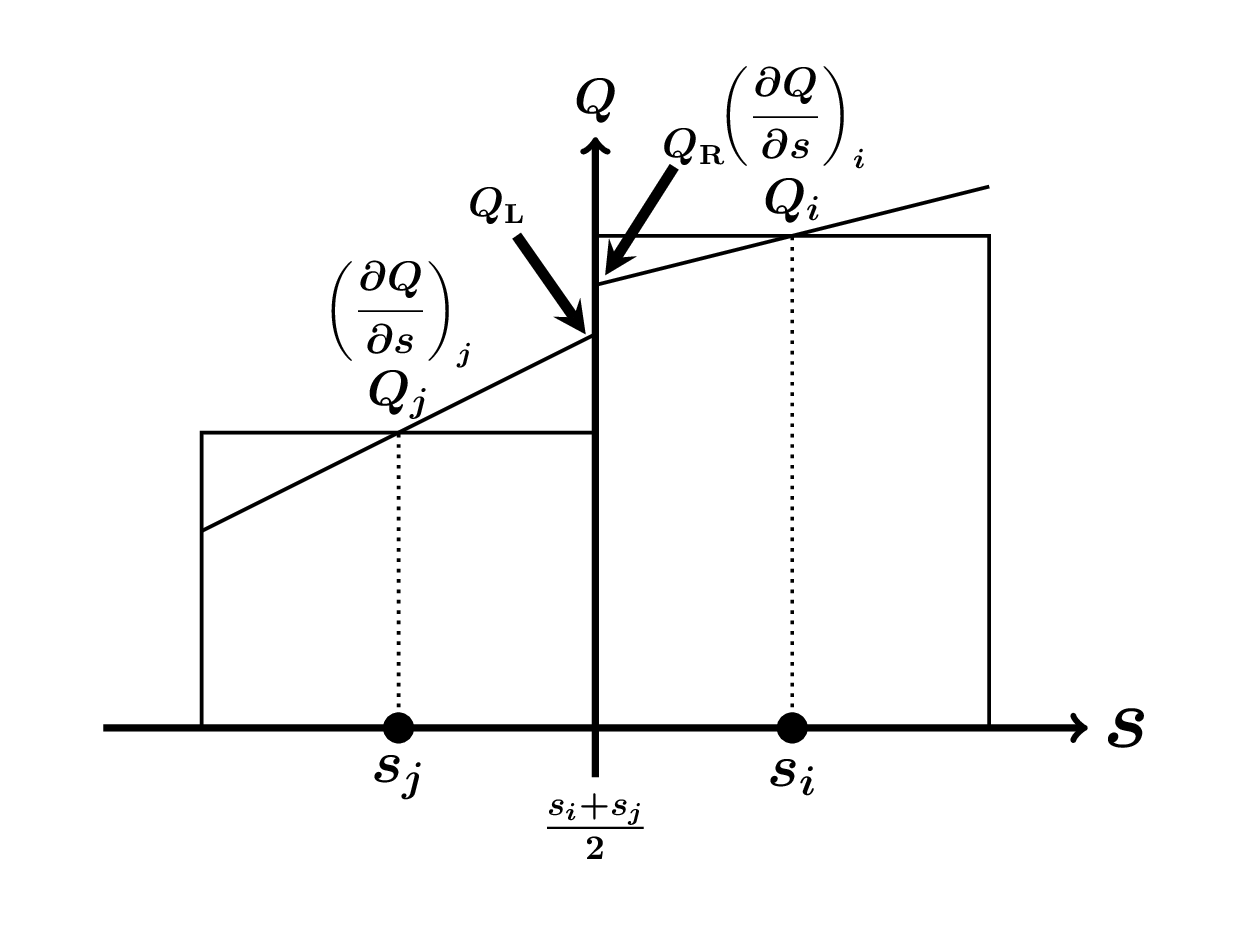}
        \end{center}
        \caption{Schematic picture of the linear interpolation}
        \label{fig:mono}
\end{figure}
To solve the RP and the MOC in the interaction between the $i$- and $j$-th particles, we need to 
evaluate physical variables $\bmath{U}_\mathrm{L}$ and $\bmath{U}_\mathrm{R}$ at $s=\left( s_i+s_j \right)/2$ 
as shown in section \ref{sec:use riemann solver}.
In the GSPMHD, a second order spatial accuracy can be achieved using a piecewise
linear interpolation of the physical variables to determine $\bmath{U}_\mathrm{L}$ and $\bmath{U}_\mathrm{R}$.
Fig. \ref{fig:mono} shows the schematic picture of the linear interpolation of a physical variable $Q$ that is 
 one of $\bmath{U}$. 
The slope of the linear interpolation of $Q$ is simply assigned by the gradient at each particle's position 
that is evaluated as 
\begin{equation}
   \left( \frac{\partial \rho}{\partial s} \right)_{i} 
   = \bmath{n}\cdot\left[\sum_k m_k \bmath{\nabla} W(\bmath{x}_i-\bmath{x}_k,h_i)
   \right],
   \label{gradient}
\end{equation}
for $Q=\rho$, otherwise
\begin{equation}
   \left( \frac{\partial Q}{\partial s} \right)_{i} 
   = \bmath{n}\cdot\left[\sum_k\frac{ m_k}{\rho_k} \left( Q_k - Q_i \right){\bmath{\nabla} }W(\bmath{x}_i-\bmath{x}_k,h_i)
   \right].
   \label{gradient}
\end{equation}
Using the gradient, the values at left and right side in the RP are given by 
\begin{equation}
        Q_\mathrm{L} = Q_j + \frac{1}{2}\Delta Q_j,\;\;\mathrm{and}\;\;\;
        Q_\mathrm{R} = Q_i - \frac{1}{2}\Delta Q_i,
\label{Qlr}
\end{equation}
respectively, where 
\begin{equation}
      \Delta Q_{i}\equiv 
        \left( \frac{\partial Q}{\partial s} \right)_{i} \Delta s_{ij},\;\;\mathrm{and}\;\;
     \Delta Q_j\equiv   \left( \frac{\partial Q}{\partial s} \right)_{j} \Delta s_{ij}
\end{equation}
If equation (\ref{gradient}) is directly used in the derivation of $Q_\mathrm{L}$ and $Q_\mathrm{R}$,
unphysical numerical oscillations arise \citep{vL79}.
In order to obtain stable second-order scheme, we need to impose a monotonicity constraints on $\Delta Q$.
In the finite-volume method, \citet{vL79} proposed several monotonicity constraints.
We apply one of them to the GSPMHD as follows
\begin{equation}
        \left(\Delta Q\right)_i^\mathrm{mono}
        = \left\{
        \begin{array}{l}
            \min\left\{2|Q_i - Q_j|,\;|\Delta Q_i|,\;
            2|\Delta Q_i'|\right\}\mathrm{sgn}(\Delta Q_i)\\
            \hspace{1cm}  \mathrm{if}\;\;\;\mathrm{sgn}(Q_i - Q_j) = \mathrm{sgn}(\Delta Q_i) 
            =\mathrm{sgn}(\Delta Q_i'),
            \\ 0\;\;\;\mathrm{otherwise,}
        \end{array}
        \right. 
        \label{mono}
\end{equation}
where $\Delta Q_i'$ satisfies 
\begin{equation}
        \Delta Q_i = \frac{(Q_i - Q_j) + \Delta Q_i'}{2}.
        \label{dq'}
\end{equation}
Here, we consider the case with $\Delta Q_i>0$ and $Q_i-Q_j>0$ as shown in Fig. \ref{fig:mono}.
The first two terms in equation (\ref{mono}), $\min\left\{2(Q_i - Q_j),\;\Delta Q_i\right\}$, ensure the 
condition of $Q_\mathrm{R}>Q_j$ (see Fig. \ref{fig:mono}).
This is the lower bound of $Q_\mathrm{R}$ for $\Delta Q_i>0$. 
In the finite-volume method, the upper bound of $Q_\mathrm{R}$ is determined by $Q$ for $s>s_i$.
On the other hand, in the SPH method, we do not know the distribution of $Q$ for $s>s_i$ explicitly 
at the instance in the 
calculation of the interaction between $i$- and $j$-th particles. However, it can be estimated by the fact that 
$\partial Q_i/\partial s$  is calculated by using all particles for $s<s_i$ and $s>s_i$, suggesting that  
$\partial Q_i/\partial s$ can be regarded as the average gradient around $\bmath{x}_i$.
If the gradient in $s_j<s<s_i$ is simply approximated by $(Q_i-Q_j)/\Delta s_{ij}$, 
the gradient in $s>s_i$, $\Delta Q'/\Delta s_{ij}$, can be guessed by equation (\ref{dq'}).
In this paper, we set the upper bound of $Q_\mathrm{R}$ by $\Delta Q_i'$ (see equation (\ref{mono}).

In the actual calculation, we take into account the domain of dependence as follows,
\begin{equation}
        Q_\mathrm{L} = Q_j + \frac{\Delta Q_j}{2} \left( 1- \frac{C_j \Delta t}{\Delta s_{ij}} \right),
        \;\;\mathrm{and}\;\;
        Q_\mathrm{R} = Q_i - \frac{\Delta Q_i}{2} \left( 1- \frac{C_i \Delta t}{\Delta s_{ij}} \right).
\end{equation}
In the RP, we use the monotonicity constraint with respect to $\rho$, $\bmath{v}\cdot \bmath{n}$, 
$P_\mathrm{t}$, $B_\perp^2$.
In the MOC, the monotonicity constraint is used with respect to the Riemann invariants $\Delta \bmath{J}_{\pm,i}$ 
(see equation (\ref{J+-})). 
Using them, we derive $\Delta \bmath{B}_{\perp i}$ and $\Delta \bmath{v}_{\perp, i}$.
Equation (\ref{mono}) suppresses numerical oscillations reasonably well. 
However, many improvements could still be made to the monotonicity constraint in the SPH method.

\end{document}